\documentclass{aa}

\usepackage{graphicx}
\usepackage{txfonts}
\usepackage{hyperref}

\begin{document} 

   \title{Strong spiral arms drive secular growth of pseudo bulges in disk galaxies}

   \author{Si-Yue Yu\inst{\ref{mpifr}}
          \and Dewang Xu\inst{\ref{kiaa}}\fnmsep\inst{\ref{pku}}
          \and Luis C. Ho\inst{\ref{kiaa}}\fnmsep\inst{\ref{pku}}
          \and Jing Wang\inst{\ref{kiaa}} \fnmsep\inst{\ref{pku}}
          \and Wei-Bo Kao\inst{\ref{pku}}
          }

   \institute{Max-Planck-Institut für Radioastronomie, Auf dem Hügel 69, 53121 Bonn, Germany\\
  			\email{syu@mpifr-bonn.mpg.de}\label{mpifr}
         \and
             Kavli Institute for Astronomy and Astrophysics, Peking University, Beijing 100871, China\label{kiaa}
         \and
             Department of Astronomy, School of Physics, Peking University, Beijing 100871, China\label{pku}
             }

\abstract{
Spiral-driven instabilities may drive gas inflow to enhance central star formation in disk galaxies.  We investigate this hypothesis using the Sloan Digital Sky Survey (SDSS) in a sample of 2779 nearby unbarred star-forming main-sequence spiral galaxies. The strength of spiral arms is quantified by their average Fourier amplitude relative to the axisymmetric disk.  The star formation properties in the central 1--3\,kpc region were derived from the SDSS spectra.  We show that galaxies with stronger spiral arms not only tend to have more intense central specific star formation rate (sSFR), larger Balmer absorption line index, and lower 4000-\AA\ break strength, but also have enhanced central sSFR relative to sSFR measured for the whole galaxy. This link is independent of redshift, stellar mass, surface density, and concentration. There is a lack of evidence for strong spiral arms being associated with a significant fraction of starburst or post-starburst galaxies, implying that the spiral-induced central star formation is likely continuous rather than bursty. We also show that stronger spiral arms tend to have an increasing fraction of pseudo bulges, a relatively unchanged fraction of star-forming classical bulges, and a decreasing fraction of quenched classical bulges. Moreover, the concentration of galaxies hosting pseudo bulges mildly increases with stronger spiral arms, implying that spirals help pseudo bulges grow.  The connection between spirals and the bulge type is partly attributed to the suppression of spirals by classical bulges and partly to the enhanced central star formation driven by spirals. We explain our results in a picture where spiral arms transport cold gas inward to trigger continuous central star formation, which facilitates the build-up of pseudo bulges. Spiral arms thus play a role in the secular evolution of disk galaxies.
}

\maketitle

\section{Introduction}

Secular evolution describes the slow rearrangement of energy and mass resulting from interactions facilitated by non-axisymmetric galaxy structures \citep{Combes1981,  Kormendy1982, Pfenniger1990, Sellwood1993, Kormendy2004}.  Secular processes dominate the evolution of galaxies in the nearby universe, while violent processes, such as major mergers at high redshift, are less common \citep[e.g.,][]{Bertone2009, Duncan2019, Montero2019}.  One of the most important secular processes, triggered by disk instability, is to drive gas to the galaxy central regions and enhance the central star formation, leading to the growth of central pseudo bulges \citep{Kormendy2004, Athanassoula2005}.  Spiral structure, a generic feature in disk galaxies, may play a role.

There are approximately $60$\% of nearby spiral galaxies hosting a bar \citep[e.g.,][]{Aguerri2009, Li2011}. The role of bars in secular evolution has been widely explored, which may give us hints about the effect of spirals. A bar imposes a non-axisymmetric potential on the disk to generate a gravitational torque which drives gas flow toward the galaxy center along the bar dust lanes \citep{Athanassoula1992b,  Regan1999, Fragkoudi2016}. Consistent with bar-driven gas transport, barred galaxies are found to have more centrally concentrated molecular gas distribution than unbarred galaxies \citep{Sakamoto1999b, Sheth2005, Kuno2007, Komugi2008}. The degree of gas concentration correlates with bar strength \citep{Kuno2007}. The inflow of gas leads to enhanced central star formation \citep[e.g.,][]{Sheth2005, Regan2006, Wang2012, Combes2014, Zhou2015, Wang2020, Simon2020}.  In particular, stronger bars tend to have more enhanced central star formation \citep{Zhou2015, Lin2017, Chown2019}, due to stronger effect in strong bars than in weak bars \citep{Regan2004}. Barred galaxies may have shorter depletion timescales measured for the whole galaxy \cite[e.g.,][]{Geron2021}. Still, spatially resolved study finds no remarkable differences in the Kennicutt-Schmidt law in the central regions of barred and unbarred galaxies \citep{Simon2021}. Strongly barred galaxies do not necessarily have enhanced central star formation rates \citep[SFRs;][]{Wang2012, Consolandi2017, Simon2020, Wang2020}. \cite{Wang2020} found that disk galaxies hosting strong bars can have both suppressed and enhanced central SFRs. Interestingly, those with enhanced central SFRs tend to connect to strong spiral arms, implying spiral arms may help to drive gas inflow. The suppressed central SFRs could result from a past starburst in which abundant gas may have existed before.  Despite the consensus that bars facilitate the pseudo bulge formation, bulges in barred galaxies do not have a different Kormendy relation \citep[][reference therein]{Kormendy2004} or different relationships between relative central surface density and other global galaxy properties than bulges in unbarred galaxies \citep{Gao2020, Luo2020}, implying that other disk structures such as spirals also participate the build-up of pseudo bulges.

Spiral-driven instabilities play a role in secular evolution.  In addition to the two well-known secular processes of heating and radial migration of stars caused by spiral arms \citep[e.g.,][]{LyndenBell1972, Athanassoula2002, Sellwood2002, Roskar2008, Sellwood2011, Sellwood2014RMP, MartinezBautista2021}, the arms could also introduce gas inflow \citep{Kalnajs1972, Roberts1972, Lubow1986, Hopkins2011, Kim2014, Baba2016, Kim2020}.  Theoretical models of quasi-static density waves predict that spiral arms can trigger large-scale shocks on cold gas as they go across the arm \citep{Roberts1969}.  The subsequent gravitational collapse induced by the shock accelerates the production of new stars. As stronger spirals trigger stronger shocks, the specific SFRs measured for the whole galaxy are found higher in galaxies with stronger arms than with weaker arms \citep{Seigar2002, Kendall2015, Yu2021}.  Studies of global gas depletion time suggest that strong spiral arms enhance star formation efficiency \citep{Yu2021}, although the efficiency does not varying much from arm to inter-arm region \citep{Foyle2010, Querejeta2021}. The large-scale spiral shocks are an efficient way to transmit angular momentum, causing gas cloud in orbital motions to move radially inward before the corotation radius, and the gravitational torque of the non-axisymmetric spiral potential acts as a secondary mechanism to drive gas inflow \citep{Kalnajs1972, Roberts1972, Lubow1986, Hopkins2011, Kim2014, Baba2016, Kim2020}. The gravitational torque of the gaseous component has an additional minor contribution of 10\% \citep{Kim2014}. Inside the corotation radius, the rate of spiral-driven gas mass inflow to the central region follows $\sim0.05-3.0$\,$M_{\odot}$yr$^{-1}$, with a more considerable inflow rate corresponding to stronger and slower-rotating arms \citep{Kim2014}.

Another picture of spiral origin, in addition to the quasi-static density waves, is the recurring spiral pattern, resulting from a recurrent cycle of groove modes \citep{Sellwood2014, Sellwood2019, SellwoodARAA}. Traditional {\it N}-body simulation of isolated disk presents recurring spiral patterns which will disappear after a few rotations due to spiral scattering \citep{Sellwood1984}. With the scattering effect much less than previously thought, recent high-resolution simulations showed that spiral arms exist much longer \citep{Fujii2011, DOnghia2013}. Moreover, longer-lived modes, which survive multiple rotations without breaking into pieces, have also been reported \citep{DOnghia2013, Sellwood2014, Sellwood2021}. Despite its transient nature, the recurring spiral patterns thus more resemble the quasi-static density waves, implying that the gas inflow driven by density waves may be applied to the recurring spiral pattern to some degree. Instructively, simulations of galactic disks subject to spiral arm perturbations of different origins suggest no apparent difference in sculpting the star-forming interstellar medium between the different models \citep{Pettitt2020}.

Compared to bars, the connection of spirals to central SFRs and the subsequent secular growth of pseudo bulges is less understood from observations. The photometric images and fiber spectroscopies available from the Sloan Digital Sky Survey \cite[SDSS;][and references therein]{York2000} provide us a good opportunity to statistically test the hypothesis that spiral-driven instabilities drive gas inward to enhance galaxy central star formation. As stronger spiral arms are more effective to drive gas inflow \citep{Kim2014}, we use the strength of spiral arms to characterize the spiral effect. The strengths of spiral arms measured based on SDSS optical {\it r}-band images are not affected significantly by emission from young massive stars \citep{Yu2021}.  There is a complex web of interdependence among spiral arm strength and other galaxy properties. Spiral arm strength correlates to galaxy mass, global SFRs, and concentration \citep{Kendall2015, YuHo2020, Yu2021}.  Meanwhile, galaxies with lower mass, surface density, and concentration have younger stellar populations in the galaxy centers averagely \citep{Kauffmann2003a, Kauffmann2003b}. We thus aim to probe the effect of spiral arms on secular evolution via the establishment of a true connection between spiral arm strength and central star formation history with effects from other galaxy parameters removed.

The paper is organized as follows. Our sample selection and methods to reduce the data are described in Section~\ref{data} and \ref{MParm}, respectively. Section~\ref{results} presents the results. The discussions are given in Section~\ref{discussion}. A summary of the main conclusions appears in Section~\ref{conclusions}.

\begin{figure*}
	\centering
	\includegraphics[width=0.7\textwidth]{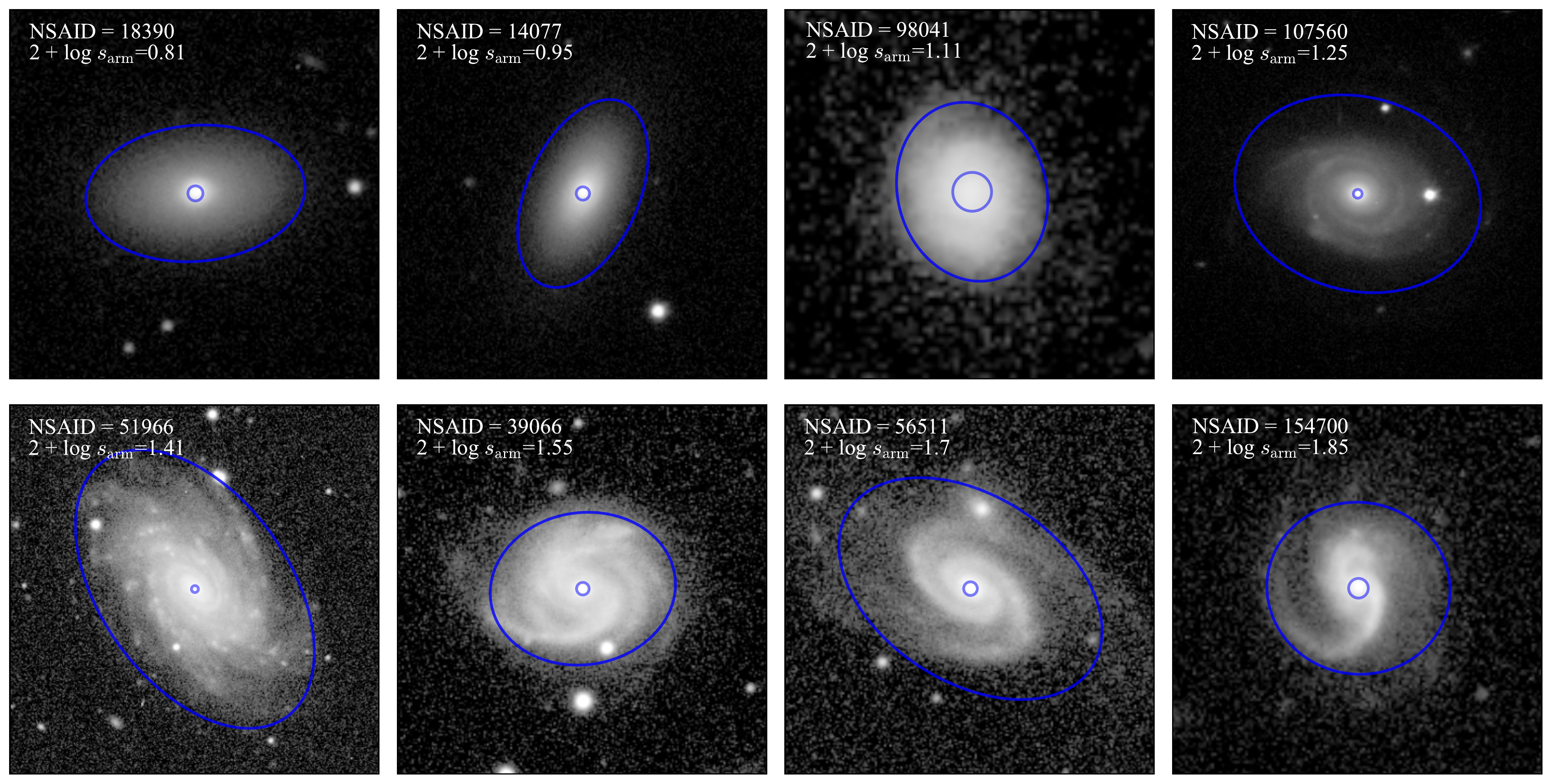}
	\caption{Example {\it r}-band galaxy images with spiral arm strength ($2+\log s_{\rm arm}$) increasing from left to right and top to bottom. The inner blue ellipse of 3\arcsec\ denotes the size of the SDSS fiber. The outer blue ellipse at a semi-major axis of $R_{90}$ illustrates the measured mean ellipticity and position angle. The ID in the NSA catalog and the arm strength is presented at the top-left corner.}
	\label{example}
\end{figure*}

\section{Sample and data}\label{data}
 
The sample studied here is derived from the NASA Sloan Atlas (NSA) \citep[][reference therein]{Blanton2011}\footnote{\url{http://www.nsatlas.org/}}, whose image background subtraction improved in the SDSS Data Release 8 \citep{Aihara2011}. To probe both central star formation properties and its enhancement relative to global star formation rates, we cross-match the NSA catalog with the MPA-JHU catalog \footnote{\url{http://www.mpa-garching.mpg.de/SDSS/DR7/}; we used the improved stellar masses from \url{http://home.strw.leidenuniv.nl/~jarle/SDSS/}} and the catalog of \cite{Salim2018}\footnote{\url{https://salims.pages.iu.edu/gswlc/}; we use GSWLC-X.}.

We select the objects with an extinction corrected {\it r} band magnitude from the NSA catalog brighter than 15 mag at redshifts $z$ less than 0.05. The magnitude limit is chosen to exclude objects with a too low signal-to-noise ratio (S$/$N). The noise would drown out the signal of structure in the outer part of galaxies if the S$/$N is sufficiently low. 
The bias caused by noise will be further corrected following the scheme in \cite{Yu2021}. The upper redshift limit is chosen to exclude galaxies whose image is degraded due to angular size shrinking caused by long distance and chosen to ensure robust quantification of spiral arms \citep{Yu2018}. We also exclude all objects at redshifts below 0.01, as the extraction of such low redshift galaxies is difficult on the basis of SDSS Atlas pipeline images. We exclude galaxies with stellar mass $\log (M_*/M_{\sun})<9$ to avoid irregular galaxies. Stellar mass ($M_*$) and star formation rate (SFR) in actively star-forming galaxies correlates with each other in a relation which is the so-called star formation main sequence \citep[MS;][]{Salim2007, Renzini2015, Saintonge2016}.  We make use of the MS defined by \cite{Saintonge2016} and select star-forming main-sequence (SFMS) galaxies by requiring $\log\,({\rm SFR_{global}}/{\rm SFR}_{\rm MS})>-2\,\sigma$, where ${\rm SFR_{MS}}$ is the SFR along the MS, $\sigma=0.4\,$dex is the scatter of the MS, and ${\rm SFR_{global}}$ is the global SFR from \cite{Salim2018}. Galaxies in quenched sequence, including ellipticals, S0s, and red spirals are excluded. Red spirals are quenched due to environmental effects such as ram pressure stripping or galaxy harassments \citep{Kormendy2012}, and their spiral arms hardly work on the secular evolution anymore. The above selection results in 7776 objects. We used the second phase of Galaxy Zoo \citep{Willett2013, Hart2016} to identify and exclude barred and edge-on galaxies. We use the redshift bias-corrected vote fraction derived in \cite{Hart2016}. We first identify edge-on galaxies as those with debiased fraction of volunteers that voted that the galaxy is edge-on $p_{\rm edgeon}\geq 0.2$, and exclude them.
Bars can enhance central star formation and will mix possible effects from spirals if they are included.  A galaxy is then classified as barred and removed if the debiased fraction of volunteers that voted that the galaxy has a bar $p_{\rm bar}\geq 0.2$ \citep{Skibba2012, Masters2012, Willett2013}.

Under our sample selection, the bar fraction $f_{\rm bar}$ reaches 53\%, which is significantly higher than $f_{\rm bar}\approx30\%$ based on the votes without redshift bias correction from Galaxy Zoo \citep{Masters2011}. We verified that this higher bar fraction is due to the fact that the votes we used were redshift-debiased \citep{Hart2016}, that our sample favors massive galaxies (median $\log M_*/M_{\odot}=10.5$), which are more likely to host a bar \citep{Erwin2018}, and that our sample only contains star-forming galaxies, which are disks, so that any possible misclassification of Es as disks, or vice versa, has been automatically ruled out. For galaxies not available in the catalog, we visually inspect the {\it r} band image to do the classification. To isolate the effect of spirals, we conduct a second inspection to exclude structures such as rings and tidal tails in case these features or processes associated with them may interfere with the inflow of gas driven by spirals. We exclude 26 galaxies severely contaminated by foreground stars or other galaxies, 25 tidally interacting or merger systems, 14 blue ellipticals, 66 galaxies with ring structures, 7 galaxies with peculiar morphology, as well as 71 more edge-on galaxies.

In case some bars may be missed in the visually inspection, we use a second method to exclude them. The profiles of ellipticity ($e$) and position angle (PA) of isophotes (Section~\ref{MParm}) are widely adopted to identify and quantify bars \citep{Athanassoula2002, Laine2002, Erwin2003, Menendez-Delmestre2007, Aguerri2009, Li2011}. The $e$ profile generally rises with increasing semi-major axis (SMA) within the bar region and then drops outside of it, and the PA correspondingly suddenly changes at the end of the bar. The difference between $e$ of an isophote and the previous one is denoted as $\Delta e$ and that for PA is denoted as $\Delta$PA. As in \cite{Menendez-Delmestre2007},  we use the criteria $\Delta e \geq 0.1$ and $|\Delta {\rm PA}| \geq 10 ^\circ$ to search for candidates hosting a bar. The PA change does not happen if the bar aligns fortuitously with the major axis of the outer disk. If there exists any isophote presenting $\Delta e \geq 0.1$ and ${\rm |PA_{iso} - PA_{disk}| \leq 20^\circ}$, the galaxy is also taken as a candidate barred system, even if $|\Delta {\rm PA}| < 10 ^\circ$. For these candidates, at each SMA that meets the criteria, we search for an isophote with a local maximum $e$ at and before this location, and set the candidate bar SMA$_{\rm bar}$, $e_{\rm bar}$, and PA$_{\rm bar}$ the same as those for this isophote. Only if there is a bar-like structure in the {\it r}-band image consistent with the candidate bar properties, is it identified as real and the galaxy classified as barred. The inspection is necessary as we found that the criteria may mistakenly identify distorted spiral arms. Finally we identify 129 barred galaxies based on the isophotal analysis and exclude them from our sample. The $f_{\rm bar}$ increases to 55\%. Short bars are generally weak \citep{Elmegreen2007}. If any bars are missed by Galaxy Zoo and our isophotal analysis, they must be short and weak. Their disks are dominated by spiral arms, so the possible missing bars may have a weak effect, but will not significantly affect our results.

In addition, 55 galaxies with disk ellipticity higher than 0.65 (Section~\ref{MParm}) are removed to avoid severe projection effects. Excluding additional 22 galaxies with S$/$N less than 2 (Section~\ref{MParm}), the selection criterion results in 2779 galaxies, the parent sample probed in this work. Our unbarred spirals have stellar mass 0.1\,dex lower than the barred galaxies that have been excluded since more massive galaxies tend to have higher bar fraction \citep{Simon2016, Erwin2018}. Although SDSS-based studies found higher frequency of bars toward more massive, gas-poor, and redder galaxies \citep[e.g.,][]{Masters2011, Masters2012}, \cite{Erwin2018} used higher quality images from the Spitzer Survey of Stellar Structure in Galaxies and showed that bars are as common in blue, gas-rich galaxies as they are in red, gas-poor galaxies. The properties of our unbarred spirals likely have no significant bias on color or gas richness compared to barred galaxies.  The defined sample is volume-limited due to the redshift and magnitude cuts. Since we are mainly interested in spiral arm characteristics of individual galaxies rather than space densities or abundances, we refrain from applying any incompleteness corrections to the results.

We utilize central SFR (${\rm SFR_{fiber}}$) and stellar mass ($M_{{\rm *,fiber}}$) within the SDSS 3\arcsec-diameter fiber from the MPA-JHU catalog. Considering the redshift ranging from 0.01 to 0.05, the 3\arcsec fiber covers a physical scale of $\sim$\,1 to 3\,kpc.  The ${\rm SFR_{fiber}}$  was estimated from the attenuation corrected H$\alpha$ luminosity for star-forming galaxies \citep{Brinchmann2004}. It traces ongoing star formation averaged over the pass $\sim$\,10\,Myr \citep{Kennicutt2012}. The $M_{{\rm *,fiber}}$ is estimated based on the dust-attenuation corrected {\it z}-band magnitude and {\it z}-band mass-to-light ratios from a Bayesian analysis \citep{Kauffmann2003a}. The central sSFR is then calculated via ${\rm sSFR_{fiber}}= {\rm SFR_{fiber}}/M_{\rm *, fiber}$.

The SFR and $M_*$ based on ultraviolet, optical, and mid-infrared photometry from the catalog of \cite{Salim2018} are used as the measures of SFR and $M_*$ for the whole galaxy (${\rm SFR_{global}}$ and $M_{*}$). The global sSFR is computed as ${\rm sSFR_{global}}={\rm SFR_{global}}/M_{*}$. A ratio of ${\rm sSFR_{fiber}}$ to ${\rm sSFR_{global}}$ is employed as a measure of relative enhancement of central star formation in \cite{Wang2012}. Following their strategy, we define the relative central enhancement of sSFR as:
\begin{equation}
  C({\rm sSFR}) = \log\,({\rm sSFR_{fiber}/sSFR_{\rm global}}).
  \label{CsSFR}
\end{equation}

Strong Balmer absorption line occurs 0.1--1\,Gyr after a burst of star formation. Its absorption line index (H$\delta_{\rm A}$) rises to maximum when hot OB stars have terminated their evolution and A stars are dominated \citep{Worthey1997, Poggianti1999, Poggianti2009, Kauffmann2003a, Dressler2004}. The H$\delta_{\rm A}$ thus probes star formation rate on intermediate timescales of 0.1--1\,Gyr prior to observation.  The 4000-\AA\ break strength, D$_n(4000)$, generated by a combination of metal absorption and the lack of flux from young and hot OB stars \citep[e.g.,][]{Poggianti1997, Balogh1999, Kauffmann2003a}, traces the current luminosity-weighted mean stellar age. In the case of an instantaneous, solar-metallicity burst of star formation, it increases from $\sim$\,1 in young stellar populations with little or no metal absorption at the age of $\sim$\,10\,Myr to $\sim$\,2 in old population with strong metal line absorption at the age of $\sim$\,10\,Gyr \citep{Kauffmann2003a}. The D$_n(4000)$ thus probes long timescales star formation history.

It is worth emphasizing that we essentially relate the spiral arms occupying the extended disk to the central-most ($1-3$\,kpc) diameter region of the galaxy. In Figure~\ref{example}, the comparison between the SDSS fiber size, marked by the inner blue circle, and the disk size, marked by the outer blue ellipses, is illustrated. In particular, the third (${\rm NSAID}=98041$) and fifth (${\rm NSAID}=51966$) panel, respectively, present two extreme cases with a large and small ratio of fiber size to disk size.

\begin{figure*}
	\centering
	\includegraphics[width=0.7\textwidth]{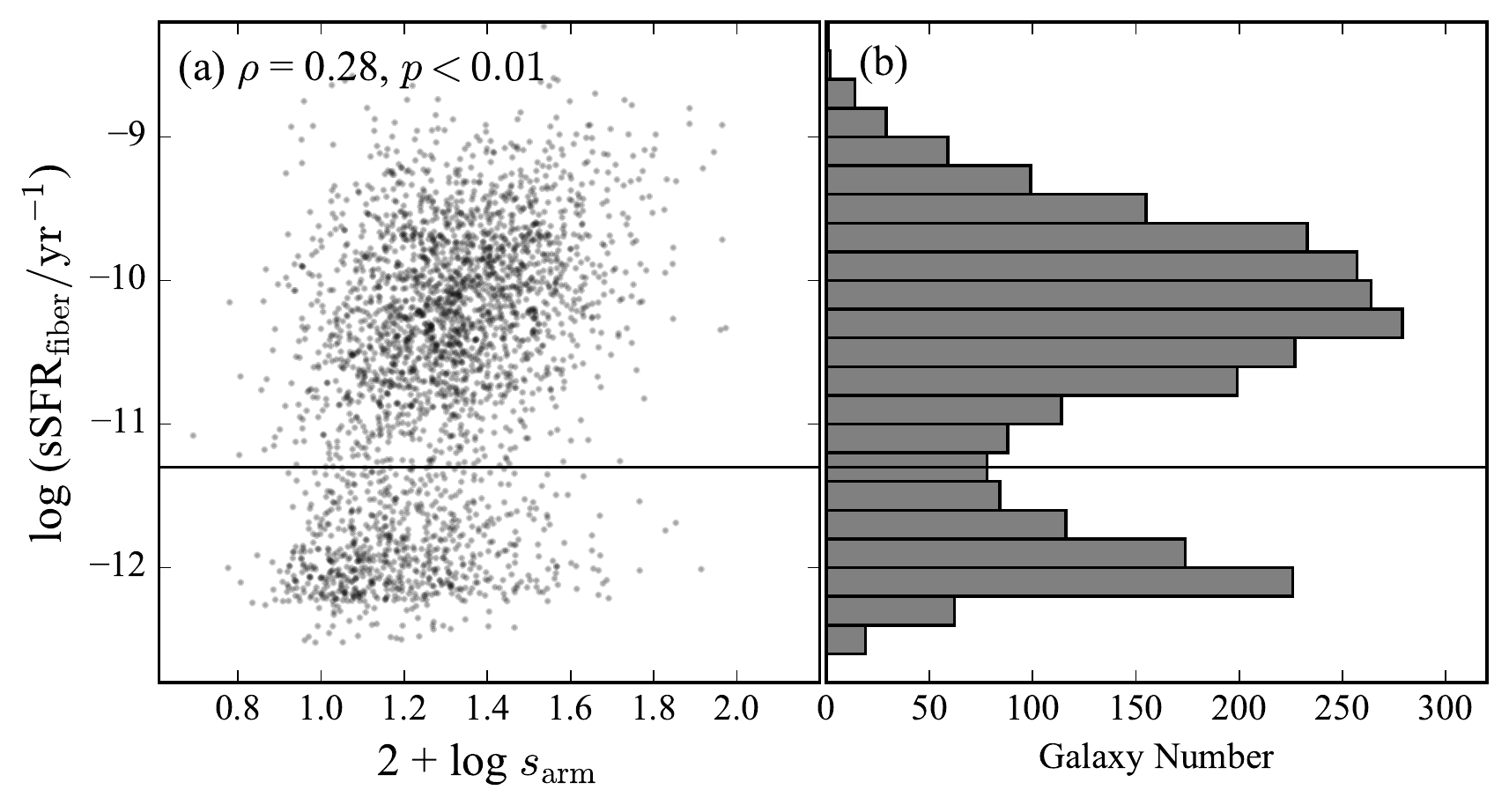}
	\caption{Dependence of central specific star formation rate within the SDSS fiber (sSFR$_{\rm fiber}$) on the strength of spiral arms ($2+\log s_{\rm arm}$) in panel (a) and bimodal number distribution of sSFR in panel (b). The solid horizon line ($\log {\rm sSFR}_{\rm fiber}=-11.3$) marks the valley of the number distribution, separating the sample into centrally star-forming SFMS galaxies and centrally quenched SFMS galaxies. The Pearson correlation coefficient between sSFR$_{\rm fiber}$ and $2+\log s_{\rm arm}$ is denoted at the top.}
	\label{fib_s}
\end{figure*}

\begin{figure*}
	\centering
	\includegraphics[width=0.8\textwidth]{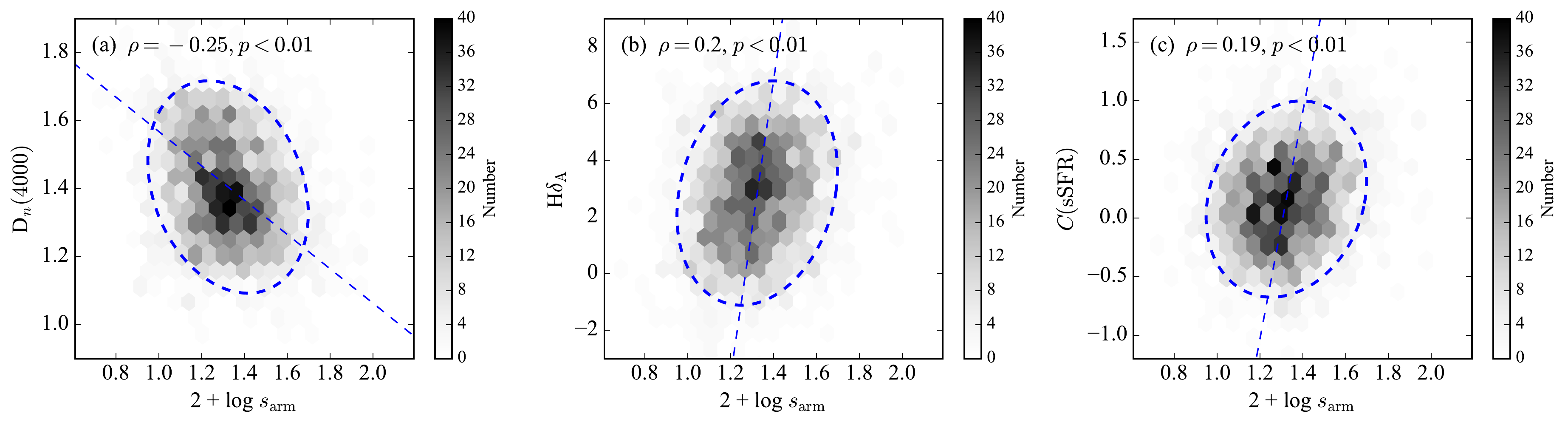}
	\caption{Dependence of (a) D$_n(4000)$, (b) H$\delta_{\rm A}$, and (c) ratio of central sSFR to sSFR measured for the whole galaxy ($C({\rm sSFR})\equiv \log\,({\rm sSFR_{fiber}/sSFR_{\rm global}})$) on spiral arm strength ($2+\log s_{\rm arm}$). The bin color encodes the number of galaxies in each bin, scaled by the colorbar next to the panel. The Pearson correlation coefficient and the corresponding $p$ value are denoted at the top. The confidence ellipses are obtained using the PCA technique described in the text, and they contain approximately 95\% of the data. As these ellipses could be distorted due to the different dynamical range of the $x$ and $y$ axis, a best-fitted straight line indicating the direction of main principal component is plotted. 
}
	\label{longsf}
\end{figure*}

\section{Morphological parameters}\label{MParm}
Our analysis to quantify spiral arm structure is based on the {\it r} band cutout images from the NSA. For each galaxy, we first generate a mask of foreground stars and background or nearby galaxies, by using both the automatic code {\tt Photutils} photometry package\footnote{http://photutils.readthedocs.io} and a manually-built mask to exclude stars inside the galaxy that {\tt Photutils} may have missed.  Instead of using the sky background-subtracted image from the NSA directly, we calculate the residual background by averaging the flux over the region in which the intensity profile becomes flat and subtract this value from the NSA {\it r}-band cutout images to produce the sky-subtracted image used in this work. We then use the {\tt IRAF} task {\tt ellipse} to obtain profiles of ellipticity ($e$) and position angle (PA) with an exponential step of 0.1.

We adopt two methods to estimate the average $e$ and PA of the disk \citep{Yu2018, Yu2019, YuHo2020}.  A combination of wrong $e$ and PA will make the round disk resemble an oval on its face-on viewing angle. The first method is based on the radial profiles of $e$ and PA. The adopted average $e$ and PA  for a galaxy is set to the average value over the region where the disk component dominates. This method fails if the disk hosts two prominent spiral arms that extend to the outskirts of the galaxy, resulting in continuously varying $e$ and PA without convergence.  In this case, we employ the alternative method, which performs a two-dimensional Fourier transformation of the disk component for a number of groups of $e$ and PA, and then search for a set of parameters that minimize the real part of the $m$\,=\,2 Fourier spectra at a radial wavenumber of zero \citep[see][for more details]{Grosbol2004, YuHo2020}. We apply the two methods to all galaxies and determine the optimal $e$ and PA for each galaxy by giving preference to that which yields a rounder image or spiral arms more closing to a logarithmic shape in the image of their face-on viewing angle.  Elliptical apertures with the derived average $e$ and PA are applied to calculate the $R_{20}$, $R_{50}$ (effective radius), $R_{80}$, and $R_{90}$, which encloses 20\%, 50\%, 80\%, and 90\% of the total flux, respectively. 96\% of galaxies have spatially resolved $R_{20}$ ($R_{20}>1\farcs4$). 20\%, 50\%, and 80\% of galaxies have the SDSS fiber radius less than 5\%, 8\%, and 10\% of the $R_{90}$, respectively. The measured $e$, PA, and $R_{90}$ for eight example galaxies are illustrated by the outer ellipses in Figure~\ref{example}. The galaxy light concentration is defined as $C= 5\times\log(R_{80}/R_{20})$ \citep{Conselice2003}. The stellar surface density is then derived via $\log\mu_*=\log (M_*/\pi R^2_{50})$.

The strength of spiral arms is one of the most fundamental properties of spirals.  The two most widely adopted methods to quantify the arm strength are (1) to compute the amplitude of Fourier components relative to the axisymmetric disk \citep{Elmegreen1989, Laurikainen2004, Elmegreen2011, Rix1995, Grosbol2004, Durbala2009, Baba2015, Kendall2011, Kendall2015, Yu2018, YuHo2020}, (2) to choose the arm and interarm region and compare their surface brightness \citep{Elmegreen1985, Buta2009, Salo2010, Elmegreen2011, Bittner2017}. We make use of the Fourier analysis due to its automatic feature with high efficiency. There are 856 galaxies overlapped with the sample in \cite{YuHo2020}, and the measurements for these galaxies are directly acquired from their work. For the rest of the galaxies, we follow the procedure described in \cite{Yu2018} and \cite{YuHo2020} to quantify spiral arms.

We first run {\tt ellipse} with fixed $e$ and PA determined above with a linear step of $R_{90}/30$. The intensity distribution, $I(r, \theta)$, as a function of azimuthal angle ($\theta$) for an isophote at radius $r$ is  extracted and then fitted with a Fourier series following

\begin{eqnarray}
	I(r, \theta) = I_{m=0}(r) + \sum_{m=1}^{6} I_{m}(r) \cos m(\theta + \phi_{m}),
\end{eqnarray}

\noindent
where $I_m$ and $\phi_m$ are the amplitude and phase angle of the $m$th Fourier component, respectively. In particular, the $I_{m=0}$ denotes the axisymmetric disk component. The relative Fourier amplitude is defined as 

\begin{eqnarray}
	A_m(r) = \frac{I_m(r)}{I_0}.
\end{eqnarray}

\noindent
$A_1$ is a measure of galaxy lopsidedness \citep[e.g.,][]{Rix1995, Reichard2008}. $A_2$ indicates the strength of spiral arms in grand-design galaxies \citep[e.g.,][]{Grosbol2004, Elmegreen2011, Kendall2011}. $A_3$ and $A_4$ reflect the arm strength in multiple-armed or flocculent galaxies \citep{Yu2018}.  Higher-order modes are not included to avoid the influence of noise. The average relative spiral arm amplitude is thus defined as the average value of a quadratic sum of the relative amplitude of $m=2$, 3, and 4 modes: 
\begin{equation}
  s_{\rm arm}=\sqrt{A_2^2+A_3^2+A_4^2}, 
\end{equation}

\noindent
over the region occupied by spiral arms. For galaxies with small bulges ($C\leq 3.5$), the inner boundary of spiral arms is set to $0.2\,R_{90}$. In contrast, for galaxies with large bulges ($C< 3.5$), it is set to a radius where the ellipticity profile drops to $e-0.05$, corresponding to the bulge size. $\Delta e=0.05$ is generally small enough. In any case, a minimum inner boundary of $0.2\,R_{90}$ is applied especially for nearly face-on galaxies. The outer boundary is set as $R_{90}$, which encloses the majority of spiral structure in optical wavelength as shown by the blue ellipse in Figure~\ref{example}. Although the relative Fourier amplitude may variate with radius, the uncertainty in calculating arm strength from choosing radial extent is less than 10\% \citep{Yu2018}, and thus hardly affect the relationships probed with arm strength \citep{Yu2021}. As the relation between arm amplitude and star formation rate is highly nonlinear, it is suggested to use the logarithmic format of the average relative amplitude, $\log s_{\rm arm}$, as the strength of spiral arms \citep{Yu2021}. We further add a constant of 2, and it becomes $2+\log s_{\rm arm}$ to make the value greater than zero in this work.

The S$/$N is defined as the value of average pixel flux between $R_{50}$ and $R_{90}$ divided by the sky background Poisson noise. Poisson noise from the sky background causes the spiral arm strength to be systematically overestimated when the S$/$N is sufficiently low since the contribution from noise to the Fourier decomposition will become significant. \cite{Yu2021} have studied the noise-induced bias as a function of S$/$N based on SDSS {\it r} band images. We thus correct the bias in our measured spiral arm strength using the result in \cite{Yu2021}. In the rest of this work, we refer to the noise-debiased spiral arm strength simply as the spiral arm strength.

We quantify spiral arms in the {\it r} band images. A question may arise: does the H$\alpha$ emission from star formation regions along the spiral arm in the {\it r} band cause severe overestimation in the arm strength?  It is found that the arm strength measured in the maps of emission from old stars derived based on the 3.6\,$\mu$m and 4.5\,$\mu$m flux from the Spitzer Survey of Stellar Structure in Galaxies \citep{Querejeta2015} is in good agreement with that in the {\it R} band, a bandpass close the {\it r} band \citep{Yu2021}. Comparing to the {\it i} band, where the H$\alpha$ emission is free, the {\it r}-band strength is only 3\% higher (0.015\,dex stronger in its logarithm; \citealt{Yu2021}).  These results suggest that the spiral arm strength based on {\it r} band images are not significantly affected by H$\alpha$ emission. Examples of eight galaxies with increasing spiral arm strength from left to right and top to bottom are shown in Figure~\ref{example}. 

\section{Results}\label{results}

We study relationships between spiral arm strength and central star formation history to investigate the possible impact of spiral arms on the secular evolution. Since the spiral arm strength \citep{YuHo2020, Yu2021} and central star formation history \citep{Kauffmann2003b, Brinchmann2004} are separately correlated with other galaxy structural parameters, we will perform an analysis with a control sample to isolate the spiral effect. We will use Pearson correlation coefficient to analyze these relationships. The Pearson correlation coefficient measures then strength of the linear monotonic correlation between two sets of data. In contrast, Spearman's correlation assesses monotonic relationships, regardless of whether they are linear or not. We have calculated the Spearman's correlation coefficients and found them to be virtually indistinguishable from the Pearson correlation coefficients. As the difference is small and linearity is the first-order approximation of any nonlinear relation, we use the Pearson correlation coefficient throughout this work.

\subsection{Dependence on central star formation properties}

Figure~\ref{fib_s}(a) presents the correlation between spiral arm strength and ${\rm sSFR_{fiber}}$. There is a clear bimodal distribution in ${\rm sSFR_{fiber}}$ with two peaks at $\sim$\,12 and $\sim$\,$-11$, although these galaxies are star-forming as a whole.  We illustrate the distribution in Figure~\ref{fib_s}(b). It lacks galaxies with ${\rm sSFR_{fiber}}$ below $-12.5$ due to the detection limit of central sSFR derived from the SDSS spectra \citep{Brinchmann2004}. The lower-value peak in the bimodal distribution suggests that these galaxy centers are quenched \citep[e.g.,][]{Kauffmann2003a, Brinchmann2004, Luo2020}, which may be caused by inside-out quenching via AGN feedback or morphological/gravitational quenching \citep{Martig2009}. AGN feedback transfers radiation to the surrounding gas to suppress gas accretion \citep{Matteo2005} or kinetic energy and momentum to cause expulsion of gas \citep{Croton2006}.  Morphological/gravitational quenching proposes that the growth of stellar spheroids stabilizes the gaseous disk to prevent the formation of bound, star-forming gas clumps \citep{Martig2009}. 25\% of SFMS galaxies are centrally quenched in our sample. In order to avoid any possible influence of these quench processes on the spiral effect investigated in this work, we separate our sample into two subsamples, 2056 centrally star-forming galaxies and 723 centrally quenched galaxies, by the valley (${\rm \log (sSFR_{fiber}/yr^{-1})} = -11.3$) of the bimodal distribution, marked by the horizon solid line in Figure~\ref{fib_s}.  We focus on the centrally star-forming galaxies in the rest of this section and come back to the centrally quenched galaxies to discuss bulge types in Section~4. A moderate trend of increasing ${\rm \log sSFR_{fiber}}$ with stronger spiral arms occurs, with a Pearson correlation coefficient $\rho=0.28$ and a $p$ value less than 0.01 (row [1] in Table~\ref{tb1}). We consider correlations with coefficients above 0.4 to be relatively strong. Correlations with coefficients between 0.2 and 0.4 are considered moderate. Correlations with coefficients below 0.2 are considered weak.

Figure~\ref{longsf}(a) presents the connection between spiral arm strength and D$_n(4000)$. The D$_n(4000)$ parameter is extracted from the SDSS 3''-diameter spectra of the galaxy center. As the data points are too crowded to evaluate the data distribution, we do not draw a scatter plot but bin the data and color the bins according to the number of galaxies in each bin. Galaxies with stronger spiral arms tend to have lower D$_n(4000)$, with a Pearson correlation coefficient $\rho=-0.25$ and a $p$ value less than 0.01 (row [6] in Table~\ref{tb1}). The D$_n(4000)$ for strong spiral arms could reach $\sim$\,$1.1$, corresponding to very young stellar populations in the centers.

In each panels of Figure~\ref{longsf}, the 2\,$\sigma$ confidence ellipse approximately marks the regions containing $\sim$\,95\% of the data. The ellipses are oriented in the direction of maximal variance of the data points, which is denoted by the dashed straight line. These ellipses are obtained using principal component analysis (PCA), which describes the data using a new set of orthonormal bases, the principal components, that successively maximize variance. The orientation of the ellipse and the slope of the straight lines are determined by the direction of the eigenvector with the largest eigenvalue, along which the data has maximum variance in the diagram. The semi-major and semi-minor axis of the ellipse is the root of the larger eigenvalue and smaller eigenvalue, respectively, of the covariance matrix of the data. Therefore, the straight line is the best-fitted linear relation between the two parameters, obtained using PCA.

Figure~\ref{longsf}(b) plots the H$\delta_{\rm A}$ as a function of spiral arm strength. The H$\delta_{\rm A}$ parameter is obtained from the SDSS 3\arcsec-diameter spectra of the galaxy center. There is a correlation, with H$\delta_{\rm A}$ rising from near $-1$ for weak arms to 7 for strong arms. The Pearson correlation coefficient gives $\rho=0.2$ with a $p$ value less than 0.01 (row [11] in Table~\ref{tb1}). The H$\delta_{\rm A}$ of a few galaxies can even reach a value higher than 7, which can only originate $\sim$\,0.1 to 1 Gyr after a star formation burst. However, the connection between H$\delta_{\rm A}$ and arm strength is  only moderate, indicated by the low correlation coefficient, so that the very high H$\delta_{\rm A}$ may hardly be attributed to the spiral effect.

\cite{Yu2021} report a correlation between spiral arm strength and global sSFR \citep[see also][]{Seigar2002, Kendall2015}, which is likely due to an intertwining process that shock of spiral arms triggers global star formation of the cold gas reservoir, which in turn maintains the arms through gas damping. The $C({\rm sSFR})$ (Equation~\ref{CsSFR}), a ratio of fiber sSFR to global sSFR, reflects relative enhancement of central star formation \citep{Wang2012, Wang2020}.  To test if the correlation between spiral arm strength and central star formation is simply driven by the global star formation, we plot the $C({\rm sSFR})$ as a function of arm strength in Figure~\ref{longsf}(c). Likewise, we detect a positive trend that galaxies with stronger spiral arms trend to have higher $C({\rm sSFR})$, namely more intense central star formation relative to that measured for the whole galaxy (Pearson correlation coefficient $\rho=0.19$ with $p<0.01$; row [16] in Table~\ref{tb1}). Therefore, the spiral arm strength positively correlates with both short-, long-timescale central SFRs and the relative SFRs enhancement, with Pearson correlation coefficients $\rho$ ranging from 0.19 to 0.28 and $p$ values $<0.01$, suggesting that these relationships are moderate but statistically significant. Our results suggest that spiral arms may play a role in enhancing central star formation.

\subsection{Analysis with a control sample}

We have shown above a dependence of central star formation history on spiral arm strength. However, both the spiral arms and central star formation history correlate with other galaxy structural parameters.  \cite{YuHo2020} showed that spiral arms become weaker in earlier-type, more centrally concentrated galaxies. For a given concentration, spiral arms are stronger in more massive galaxies. \cite{Kauffmann2003a, Kauffmann2003b} showed that galaxy centers are averagely younger in galaxies with lower mass, surface density, and concentration. These results may lead to an indirect relationship without causality between spiral arm strength and central star formation history. For the sake of establishing a true causal relationship, it is essential to demonstrate that this relationship does not result from the known correlations for stellar mass ($\log M_{*}$), stellar surface density ($\log \mu_*$), and concentration ($C$).

One possible concern with using the fixed 3\arcsec-diameter aperture of SDSS fiber is that this size corresponds to a larger physical scale for galaxies at high redshift than those at low redshift. In particular, the 3\arcsec\ indicates $\sim$\,3\,kpc at $z=0.05$, while it indicates $\sim$\,0.6\,kpc at $z=0.01$. The fiber could include disk components of galaxies at high redshift and may cause the central star formation history to be biased toward the younger population compared with galaxies at low redshift. Thus the redshift effect needs to be removed when studying the spiral effect. In addition, the physical size ($\log R_{50}$) of disk galaxies can vary by 0.3--0.5\,dex at fixed stellar mass according to the mass-size relation \citep[e.g.,][]{Shen2003}, which may introduce an aperture effect similar with that induced by redshift. The third and fifth panels of Figure~\ref{example} illustrate the joint effect of redshift and size. The galaxy NSAID\,$=$\,98041 has $z$\,=\,0.042 and $R_{50}$\,=\,3.2\,kpc, and  has very large relative fiber aperture (the inner circle). In contrast, the galaxy NSAID\,$=$\,51966 has $z$\,=\,0.017 and $R_{50}$\,=\,10.9\,kpc, and  has very small relative fiber aperture. Since the size is implicitly involved in $\log \mu_*$ and $C$, the size effect actually has already been considered if $\log M_{*}$, $\log \mu_*$, and $C$ are simultaneously controlled. Nevertheless, we remove the size effect to rule out possible worries.

We perform an analysis using a control sample to isolate the effect of spiral arms. We first define a strong-armed sample with galaxies with spiral arm strengths that occupy the top 20\% of the sample of centrally star-forming SFMS galaxies  ($2+\log s_{\rm arm} \geq 1.44$). The remaining galaxies make up the temporary weak-armed sample ($2+\log s_{\rm arm} < 1.44$). For each galaxy in the strong-armed sample, we then randomly find within the temporary weak-armed sample a galaxy having almost the same $\log M_{\rm{*}}$, $\log \mu_*$, $C$, and $z$. The matching criterion follows $\lvert\Delta\log\mu_*\rvert\leq 0.1$, $\lvert\Delta\log M_{\rm{*}}\rvert\leq 0.1$, $\lvert\Delta C\rvert\leq 0.1$, $\lvert\Delta z\rvert\leq 0.005$, and $\lvert\Delta \log R_{50}\rvert\leq 0.1$. If such a galaxy is not found, the strong-armed galaxy is also removed from the strong-armed sample. These newly selected galaxies constitute the control sample, which is named as control weak-armed sample. The strong-armed sample and control weak-armed sample each has 482 objects.

\begin{figure*}
	\centering
	\includegraphics[width=0.8\textwidth]{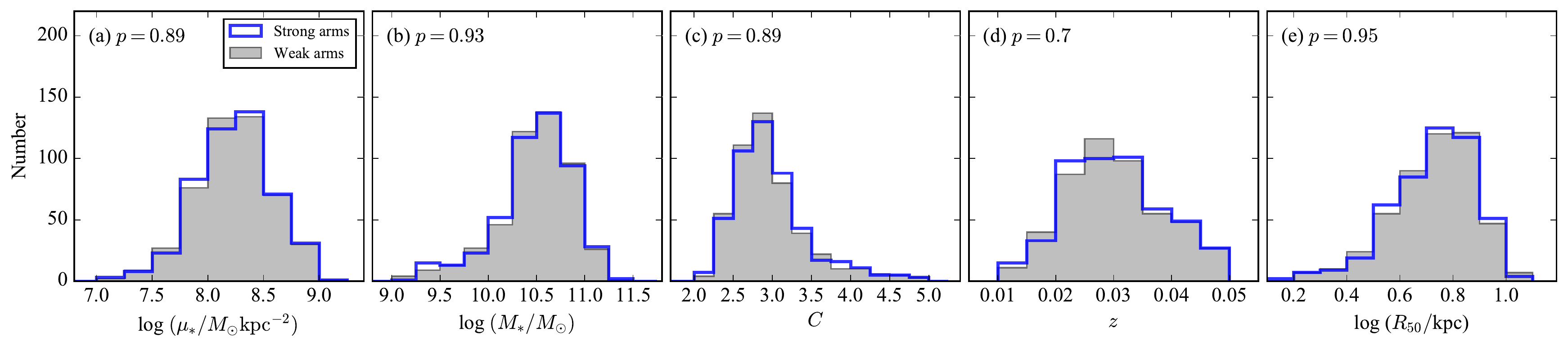}
	\caption{Number distribution of (a) stellar surface density ($\log \mu_*$), (b) stellar mass ($\log M_{\rm{*}}$), (c) concentration ($C$), (d) redshift ($z$), and (e) effective radius ($\log R_{50}$) of galaxies with strong spiral arms ($2+\log s_{\rm arm}>1.44$), marked by blue histogram, and that of the control sample with weak spiral arms, marked by gray histogram. The $p$ value of the Kolmogorov–Smirnov test is presented at the top of each panel.
}
	\label{contrl}
\end{figure*}

In Figure~\ref{contrl}, we compare the number distribution of $\log M_{*}$ in panel (a), $\log \mu_*$ in panel (b), $C$ in panel (c), $z$ in panel (d), and $\log R_{50}$ in panel (e) for the strong-armed sample, marked in blue, and the control weak-armed sample, marked in gray. For the two distributions in each panel, a two-sample Kolmogorov–Smirnov test gives $p$ values of 0.89, 0.93, 0.89, 0.7, and 0.95 respectively. Thus we cannot reject the null hypothesis that the two samples are drawn from the same parent distribution. In other words, the strong-armed sample and the control weak-armed sample are almost twins concerning $z$, $\log M_{*}$, $\log \mu_*$, and $C$, which, in turn, validates our procedure to generate the control sample.

Figure~\ref{comp}(a), (b), (c), and (d), respectively, compares the number distribution of ${\rm \log(sSFR_{\rm fiber})}$, $C({\rm sSFR})$, D$_n(4000)$, and H$\delta_{\rm A}$ for the strong-armed sample, marked in blue, and the control weak-armed sample, marked in gray. The $p$ values of Kolmogorov–Smirnov tests for the two samples and the mean difference between the two parameters are denoted at the top of each panel. The $p$ values are all less than 0.01, suggesting that we can reject the null hypothesis that the two samples are drawn from the same parent distribution. The galaxies with strong arms averagely have 0.4\,dex higher ${\rm \log(sSFR_{\rm fiber})}$, 0.2\,dex higher $C({\rm sSFR})$, $-0.1$ lower D$_n(4000)$, and 1.0 higher H$\delta_{\rm A}$ than those with weak arms, even after the $z$, $\log M_{*}$, $\log \mu_*$, and $C$ have been controlled.

\subsection{Partial correlation coefficients}

\begin{figure*}
	\centering
	\includegraphics[width=0.8\textwidth]{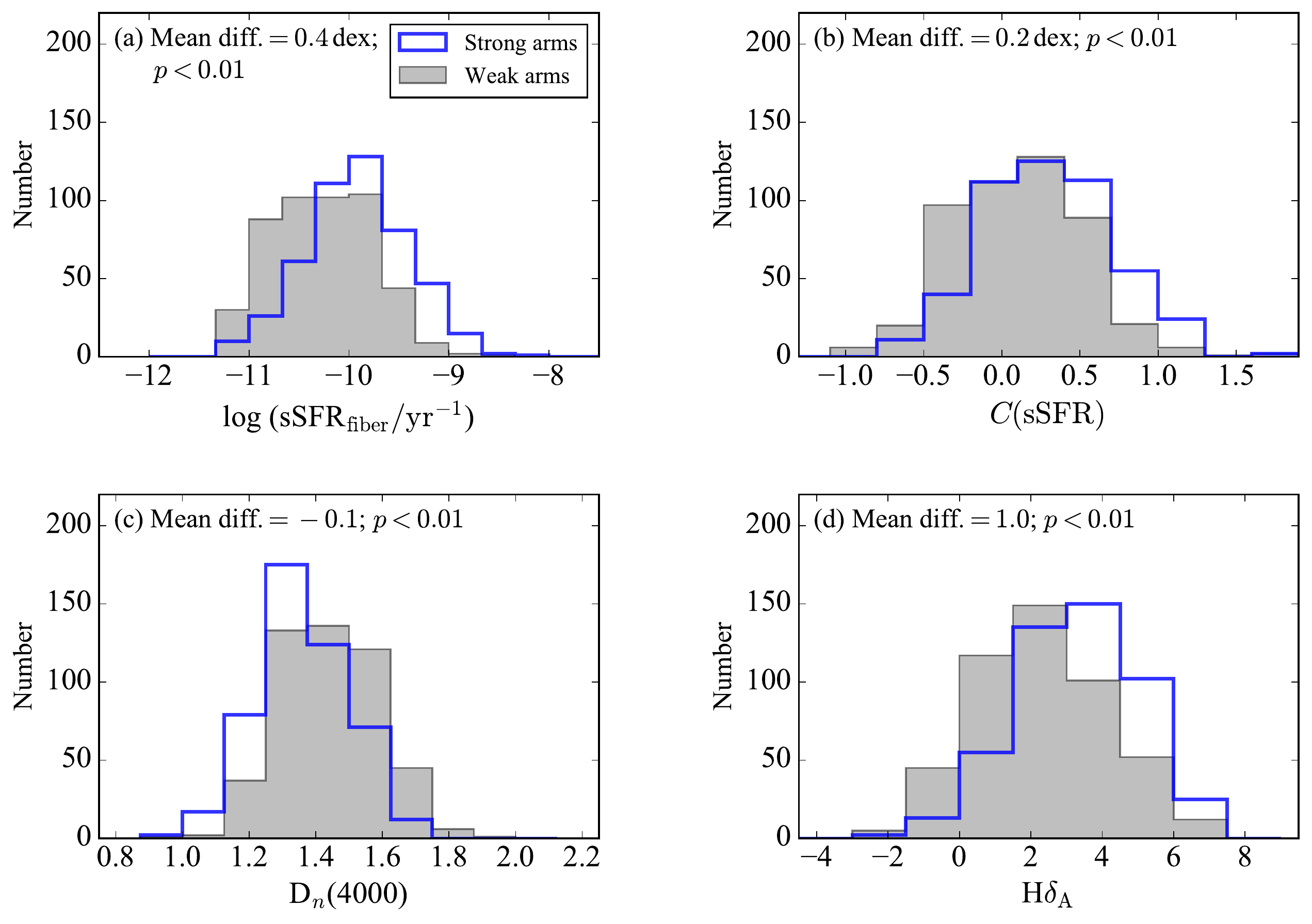}
	\caption{Number distribution of central sSFR (sSFR$_{\rm fiber}$), ratio of sSFR$_{\rm fiber}$ to sSFR measured for the whole galaxy ($C({\rm sSFR})\equiv \log\,({\rm sSFR_{fiber}/sSFR_{\rm global}})$), D$_n(4000)$, and H$\delta_{\rm A}$ for the strong-armed galaxies, marked in blue, and that for the control weak-armed sample with similar stellar mass, stellar surface density, concentration, and redshift, marked in gray. Mean difference between the two distributions and  $p$ values from the Kolmogorov–Smirnov test are presented at the top of each panel.}
	\label{comp}
\end{figure*}

The control experiment demonstrates that the dependence of central star formation on spiral arm strength is intrinsic.  The arm strength-central star formation relation may be driven to some extent or, conversely, diluted by the other parameters. In order to estimate the true strength of the relationships and to make our results statistically more robust, we compute partial correlation coefficients using python package {\tt pingouin} \citep{Vallat2018} based on inverse covariance matrix and present the results in Table~\ref{tb1}. We first separately remove the mutual dependence on $\log M_{*}$, $\log \mu_*$, and $C$ one by one ($z$ is always included to take into account for the non-physical aperture effect) to calculate the residual dependence of ${\rm \log(sSFR_{\rm fiber})}$, $C({\rm sSFR})$, D$_n(4000)$, and H$\delta_{\rm A}$ on arm strength. We did not include $\log R_{50}$ as it had been involved already in the definition of $\log \mu_*$ and $C$. The resulting partial correlation coefficients (rows [2]-[4], [7]-[9], [12]-[14], [17]-[19] in Table~\ref{tb1}) are $\sim$\,0.01 to 0.08 stronger than the original correlation coefficients (rows [1], [6], [11], [16]).  We then simultaneously remove the effects of $z$, $\log R_{50}$, $\log M_{*}$, $\log \mu_*$, and $C$, and find that the resulting partial correlation coefficients increase to 0.39, $-0.38$, 0.35, and 0.26 with $p$ values $<0.01$ (rows [5], [10], [15], [20]). These results are in agreement with the control sample analysis in Figure~\ref{comp} that neither the $z$, $\log M_{*}$, $\log \mu_*$, nor $C$ can explain the connection between arm strength and the central star formation. Furthermore, the four parameters dilute the observed relations for spiral arms.

\subsection{Spirals induce continuous central star formation}\label{csf}

We find more intense ongoing star formation rates in the centers of galaxies with stronger spiral arms, and the trend is not driven by other galaxy parameters. To understand if the spiral arms are powerful enough to trigger a central starburst, we define galaxies with ${\rm \log\,(sSFR/yr^{-1}) \geq -9}$, corresponding to a mass-doubling time of 1\,Gyr, as central starburst galaxies. We thus have 47 central starburst galaxies. The arm strengths $2+\log\,s_{\rm arm}=1.03$ and 1.64 are respectively critical values containing bottom and top 5\% of the arm strengths in the sample of 2056 centrally star-forming SFMS galaxies. The spiral arm strength of these starburst galaxies spreads over a wide dynamical range (Figure~\ref{fib_s}). 4 central starburst galaxies have very weak spiral arms ($<1.03$). In contrast, 10 central starburst galaxies have very strong spiral arms ($>1.64$). Likewise, only $\sim$10\% of very strong spiral arms are central starburst galaxies. The wide distribution of arm strength suggests that normal (non-interacting) strong spiral arm is a neither sufficient nor necessary condition to form a central starburst, although we do detect a weak trend that spiral arms in central starburst galaxies tend to be stronger.

\begin{figure*}
	\centering
	\includegraphics[width=0.9\textwidth]{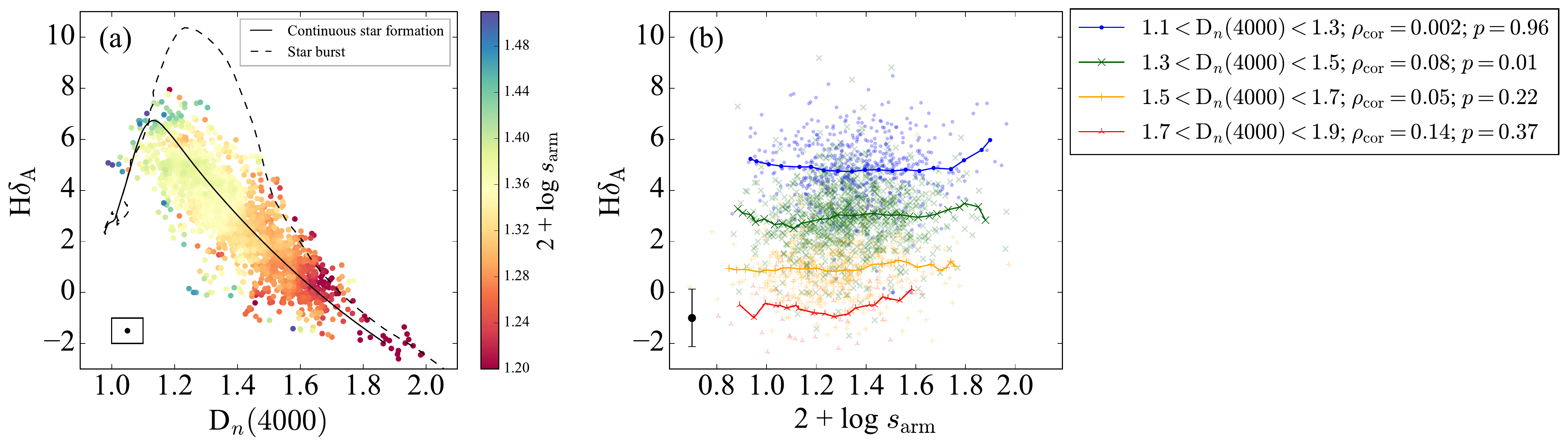}
	\caption{H$\delta_{\rm A}$ plotted as a function of D$_n(4000)$ and spiral arm strength ($2+\log s_{\rm arm}$). In Panels (a), the color associated with each data point encodes the average arm strength of surrounding galaxies with $\lvert \Delta x \rvert \leq 0.05$ and $\lvert\Delta y \rvert \leq 0.5$ (the box is illustrated in the bottom-left corner). The curves show evolution tracks obtained using the GALAXEV stellar population synthesis code \citep{bc03} with the provided simple stellar population model of a metallicity $Z=0.019$ and an \cite{Chabrier2003} initial mass function.  Dashed and solid curves are, respectively, for an instantaneous burst of star formation and continuous star formation that declines exponentially with time with a characteristic time scale of 4\,Gyr. In Panel (b), the color encodes the different narrow ranges of D$_n(4000)$. The typical scatter in H$\delta_{\rm A}$ at a given $2+\log s_{\rm arm}$ and a given D$_n(4000)$ is illustrated at the bottom-left corner. 
}
	\label{burst}
\end{figure*}

We then investigate the connection between spiral arms and their central post-starburst properties. The combination of D$_n(4000)$ and H$\delta_{\rm A}$ can be used to identify galaxies that have experienced a burst of star formation $\sim 1$\,Gyr prior to the current observation \citep{Kauffmann2003a}. Figure~\ref{burst}(a) probes H$\delta_{\rm A}$ plotted a function of D$_n(4000)$, with the color associated with each data point encodes the average arm strength of its surrounding galaxies within a box of $\lvert \Delta {\rm H}\delta_{\rm A} \rvert \leq 0.5$ and $\lvert\Delta {\rm D}_n(4000) \rvert \leq 0.05$. Consistent with Figure~\ref{longsf}, H$\delta_{\rm A}$ increases and D$_n(4000)$ decreases with stronger spiral arms. The curves in Figure~\ref{burst}(a) show evolution tracks obtained using the GALAXEV stellar population synthesis code \citep{bc03} with the provided simple stellar population model of a metallicity $Z=0.019$ and a \cite{Chabrier2003} initial mass function. Dashed and solid curves are for an instantaneous burst of star formation and continuous star formation that declines exponentially with time with a characteristic time scale of 4\,Gyr. For the continuous star formation history, H$\delta_{\rm A}$ decreases with increasing D$_n(4000)$ following a nearly linear relation. For the burst of star formation, H$\delta_{\rm A}$ quickly reach a peak before $\sim 1$\,Gyr beyond which it drops down quickly, while the D$_n(4000)$ increases gradually with time. The way that H$\delta_{\rm A}$ and D$_n(4000)$ respond differently results in a curve in D$_n(4000)$-H$\delta_{\rm A}$ diagram that H$\delta_{\rm A}$ would first increase steeply and peak at H$\delta_{\rm A}\approx 10$ with D$_n(4000)\approx 1.3$, and then decrease. If spiral arms can trigger a central burst of star formation overlapped with the existed continuous star formation, the burst will add more H$\delta_{\rm A}$ with less contribution to D$_n(4000)$ in $\sim$\,1\,Gyr. In this case, a trend that higher H$\delta_{\rm A}$ are associated with stronger spiral arms for a given D$_n(4000)$ is expected. However, there is no apparent trend of increasing H$\delta_{\rm A}$ with stronger arms for a given D$_n(4000)$ in Figure~\ref{burst}(a).

In Figure~\ref{burst}(b), we plot the H$\delta_{\rm A}$ as a function of arm strength for several bins of D$_n(4000)$. We only detect a very weak positive trend in the range of $1.3<{\rm D}_n(4000)<1.5$ with Pearson correlation coefficient $\rho=0.08$ and a $p$ value $=0.01$.  All the others have too high $p$ values and thus are not statistically significant. By removing mutual dependence on D$_n(4000)$, the partial correlation coefficient between H$\delta_{\rm A}$ and arm strength gives $\rho=0.02$ with $p$ value $=0.45$, suggesting that this residual relation does not exist. These results imply that the central star formation related to spiral arms is likely continuous instead of bursty.

\begin{figure}
	\centering
	\includegraphics[width=0.4\textwidth]{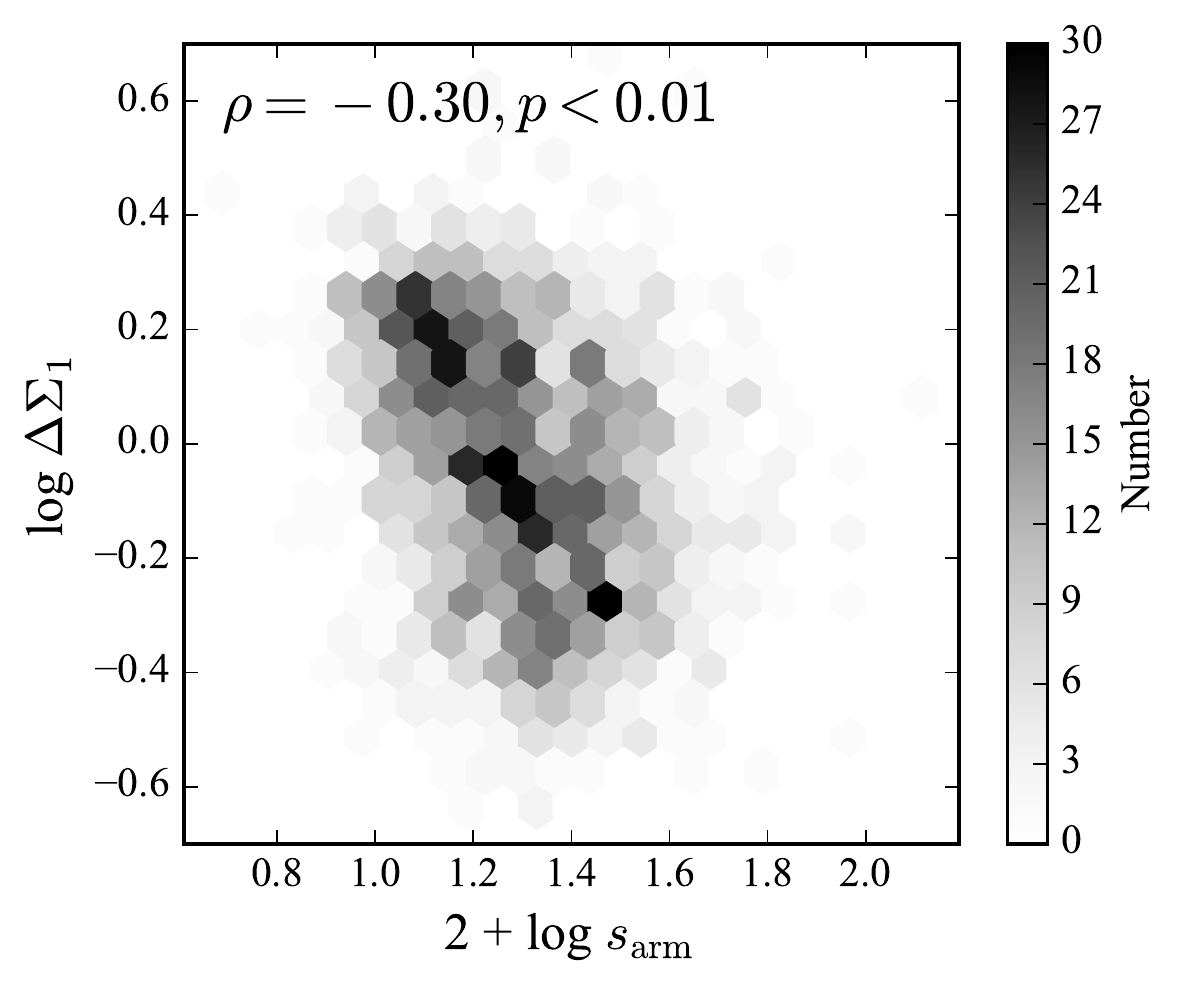}
	\caption{Dependence of relative central surface density	($\Delta\Sigma_1$) on spiral arm strength ($2+\log s_{\rm arm}$). The bin color encodes the galaxy number in each bin. $\Sigma_1$ is the stellar surface density within central 1\,kpc, while the $\log\Delta\Sigma_1$ is calculated via $\log\Delta\Sigma_1=\log\Sigma_1 + 0.275(\log M_*)^2 - 6.445\log M_* + 28.059$, which is a measure to distinguish pseudo bulge ($\log\Delta\Sigma_1<0$) and classical bulge ($\log\Delta\Sigma_1\geq 0$). The Pearson correlation coefficient is denoted at the top.}
	\label{sigma1}
\end{figure}

\begin{figure*}
	\centering
	\includegraphics[width=0.8\textwidth]{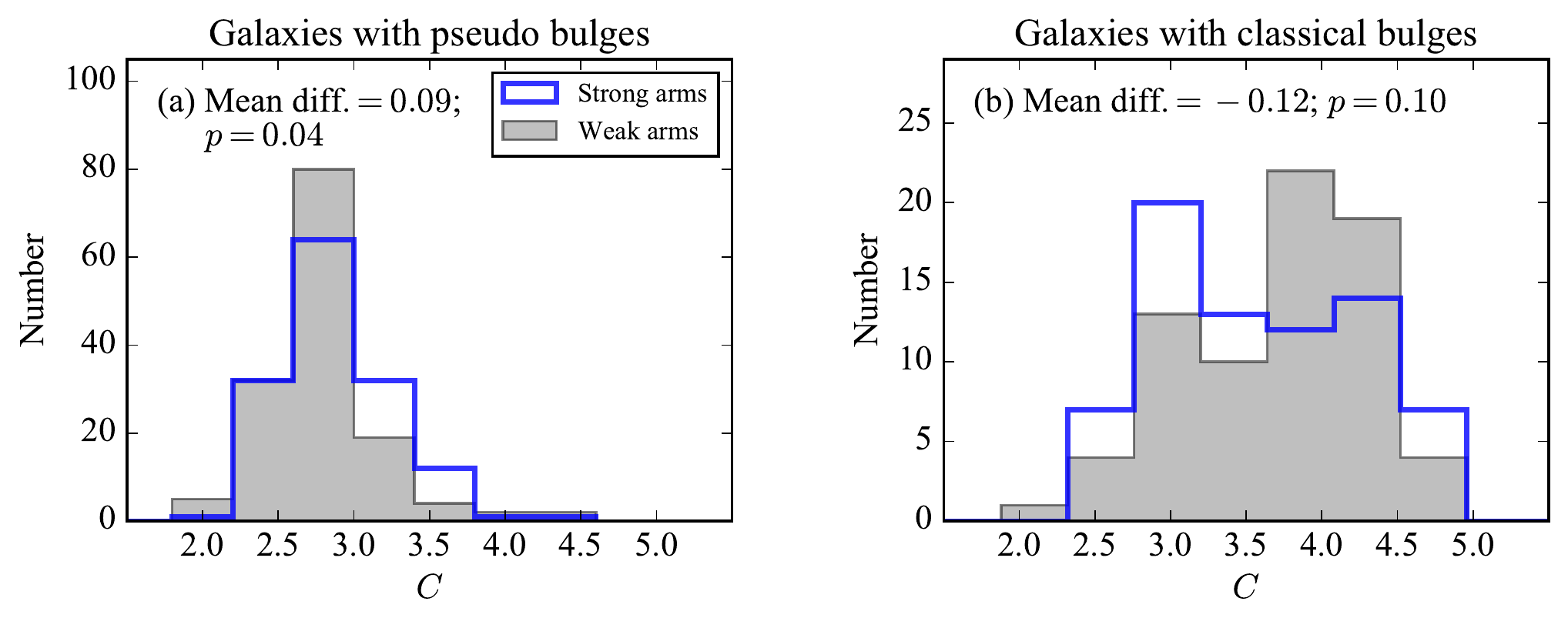}
	\caption{Number distribution of concentration index ($C$) of galaxies with pseudo bulges (a) and those with classical bulges (b). The blue and gray histogram marks the strong-armed sample and control weak-armed sample. Mean difference between the two distributions and $p$ values from the Kolmogorov–Smirnov test are presented at the top of each panel.}
	\label{CBPB}
\end{figure*}

\begin{figure}
	\centering
	\includegraphics[width=0.4\textwidth]{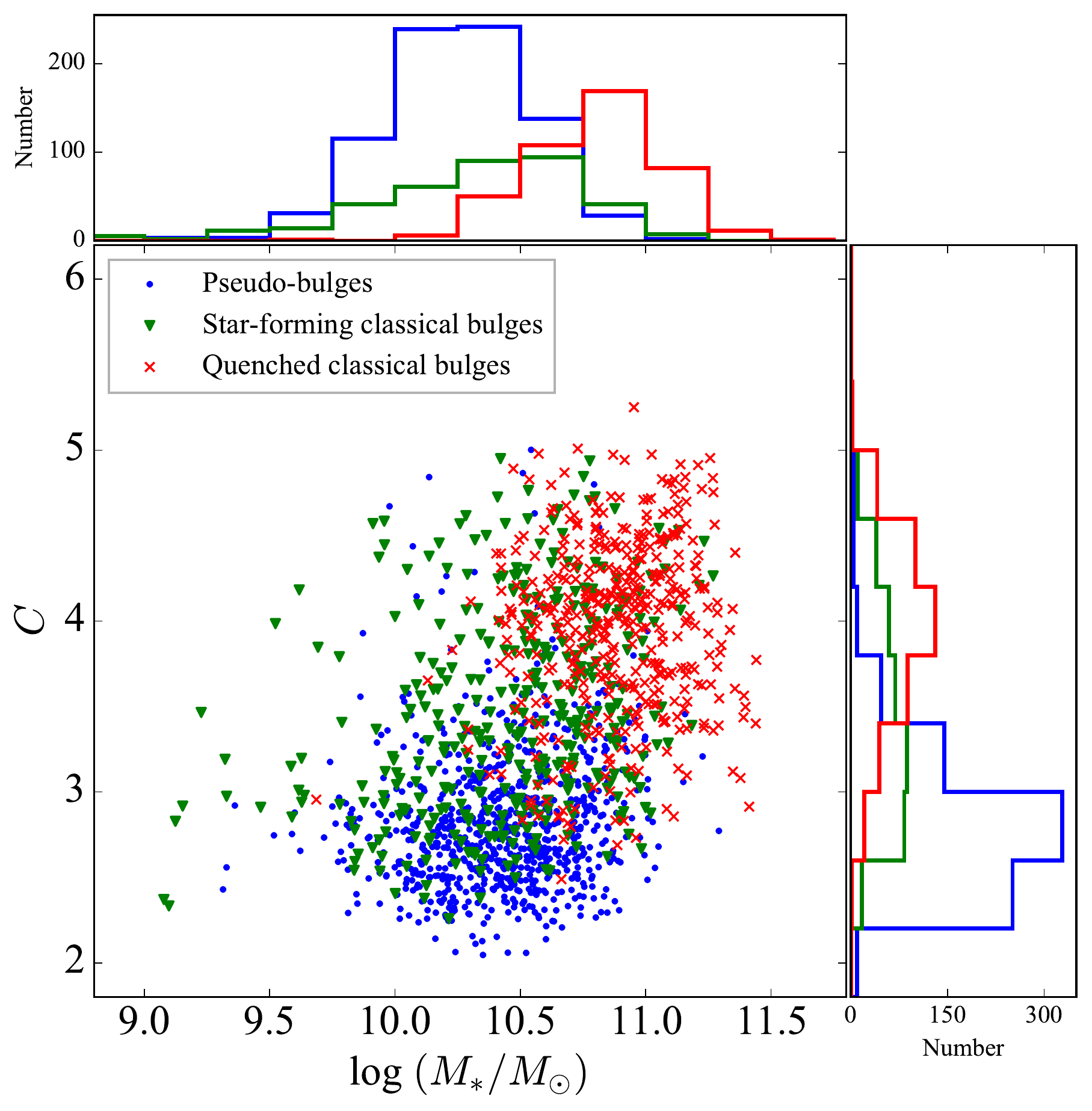}
	\caption{Concentration index ($C$) plotted as a function of galaxy stellar mass ($\log M_{*}/M_{\odot}$) for galaxies with pseudo bulges, marked in blue, star-forming classical bulges, marked in green, and quenched classical bulges, marked in red. The upper and right histogram show the number distribution of $\log M_{*}/M_{\odot}$ and $C$, respectively.
	}
	\label{mc}
\end{figure}

\begin{figure}
	\centering
	\includegraphics[width=0.4\textwidth]{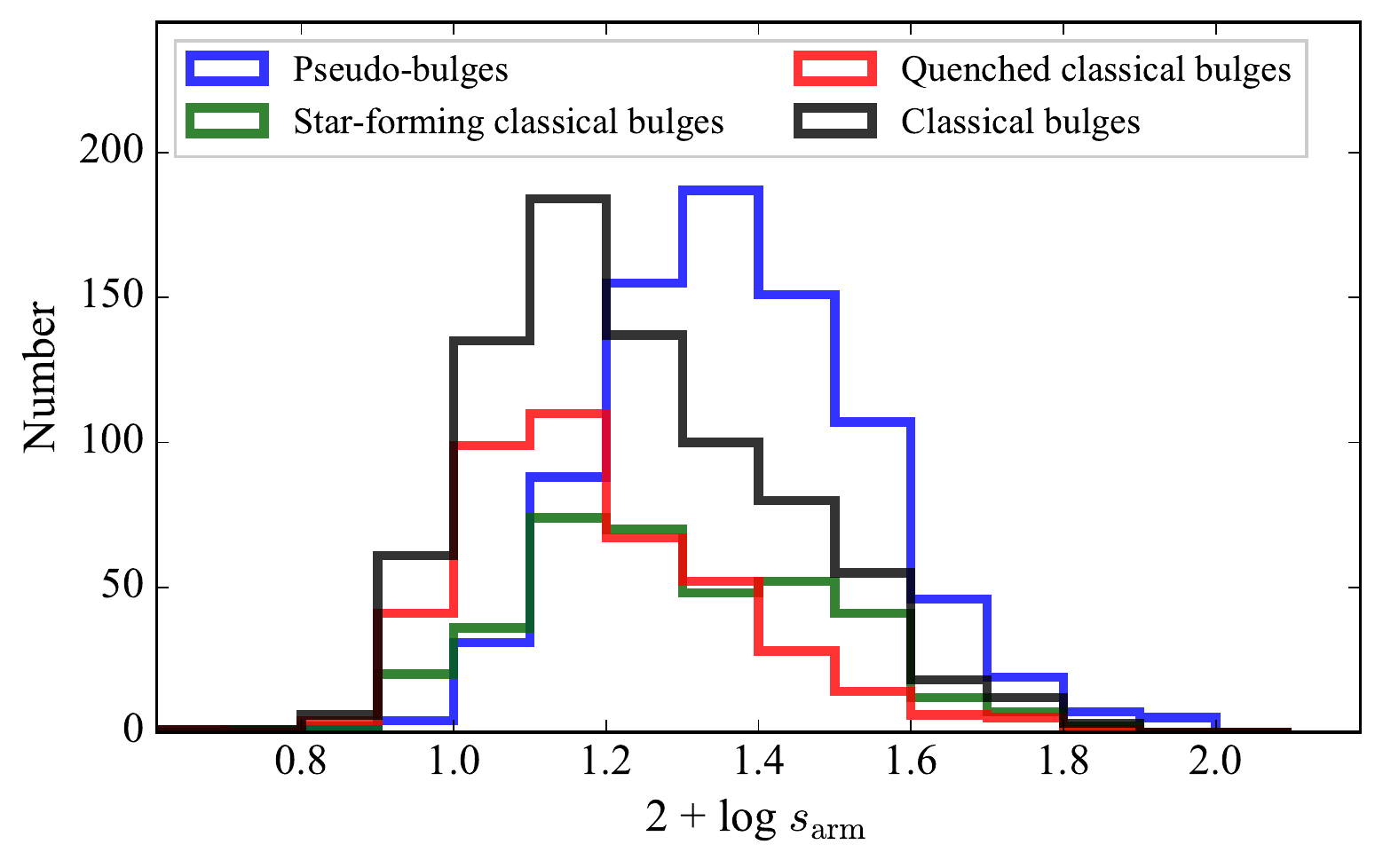}
	\caption{Number distribution of pseudo bulges (blue), classical bulges (black), star-forming classical bulges (green), and quenched classical bulges (red).}
	\label{histbul}
\end{figure}

\begin{figure}
	\centering
	\includegraphics[width=0.4\textwidth]{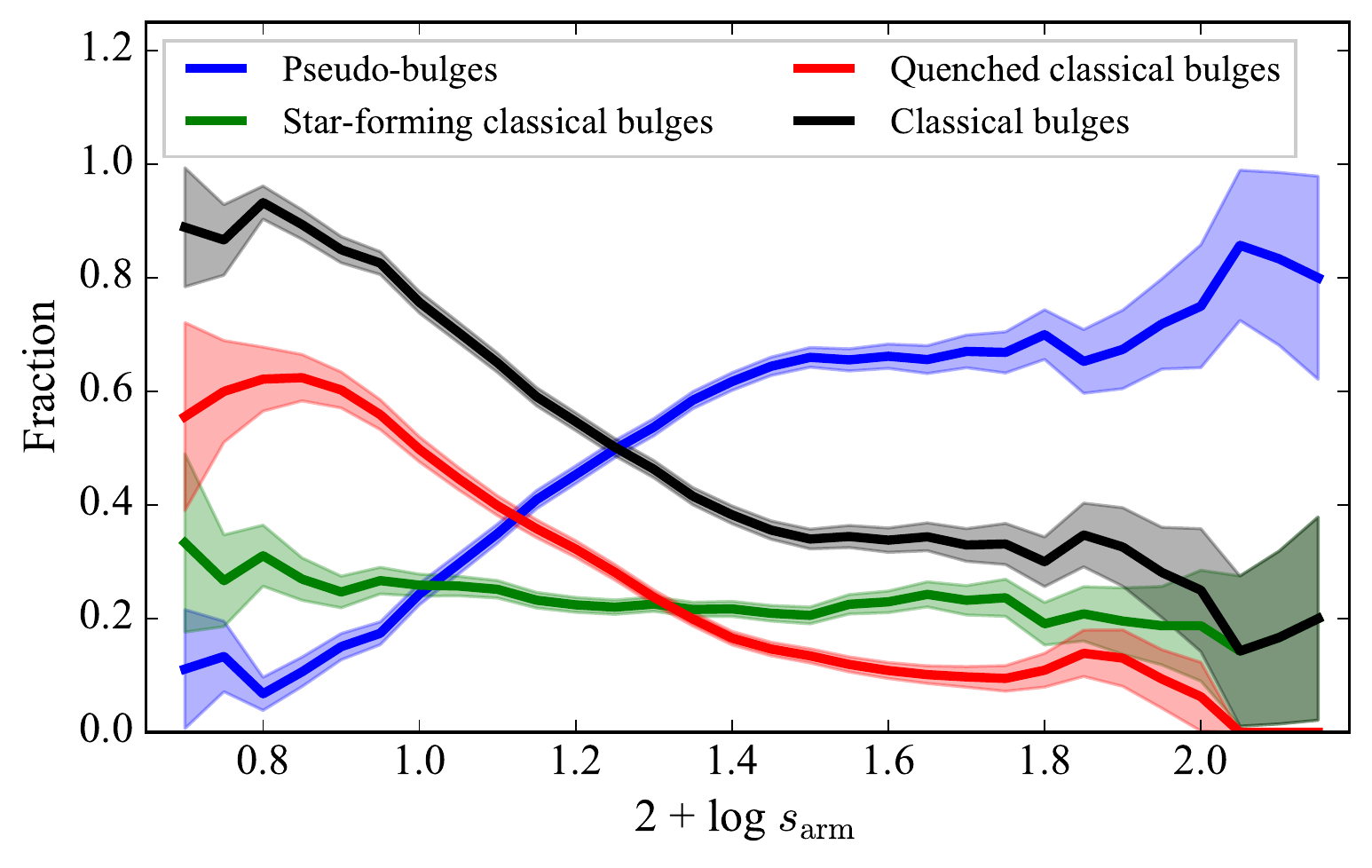}
	\caption{Fraction of galaxies hosting pseudo bulges (blue), classical bulges (black), star-forming classical bulges (green), and quenched classical bulges (red) as a function of spiral arm strength ($2+\log s_{\rm arm}$).}
	\label{frac}
\end{figure}

\begin{table*}
\caption{\label{t1}Correlation Analysis}
\renewcommand\arraystretch{1.25}
\centering
\begin{tabular}{clccccc}
\hline\hline
Row & Parameter 1 & Parameter 2 & Dependence & Partial Correlation & $p$ & Sample \\
    &        &             & Removed    & Coefficient  & \\
 (1)   & (2) & (3)         & (4)    & (5)  & (6) & (7) \\
\hline
$[1]$&$\log\,s_{\rm arm}$ &  $\log\,\rm{sSFR_{fiber}}$ & $\dots$                          & 0.28 & $<0.01$ & $\rm{sSFR_{fiber}}>10^{-11.3}\,yr^{-1}$ \\
$[2]$&$\log\,s_{\rm arm}$ &  $\log\,\rm{sSFR_{fiber}}$ & $z$, $\log M_*$                  & 0.29 & $<0.01$ & $\rm{sSFR_{fiber}}>10^{-11.3}\,yr^{-1}$\\
$[3]$&$\log\,s_{\rm arm}$ &  $\log\,\rm{sSFR_{fiber}}$ & $z$, $\log \mu_*$                & 0.33 & $<0.01$ & $\rm{sSFR_{fiber}}>10^{-11.3}\,yr^{-1}$\\
$[4]$&$\log\,s_{\rm arm}$ &  $\log\,\rm{sSFR_{fiber}}$ & $z$, $C$                         & 0.29 & $<0.01$ & $\rm{sSFR_{fiber}}>10^{-11.3}\,yr^{-1}$\\
$[5]$&$\log\,s_{\rm arm}$ &  $\log\,\rm{sSFR_{fiber}}$ & $z$, $\log R_{50}$, $\log M_*$, $\log \mu_*$, $C$ & 0.39 & $<0.01$ & $\rm{sSFR_{fiber}}>10^{-11.3}\,yr^{-1}$\\
\hline
$[6]$&$\log\,s_{\rm arm}$ & D$_n(4000)$ & $\dots$                          & $-0.25$ & $<0.01$ & $\rm{sSFR_{fiber}}>10^{-11.3}\,yr^{-1}$\\
$[7]$&$\log\,s_{\rm arm}$ & D$_n(4000)$ & $z$, $\log M_*$                  & $-0.32$ & $<0.01$ & $\rm{sSFR_{fiber}}>10^{-11.3}\,yr^{-1}$\\
$[8]$&$\log\,s_{\rm arm}$ & D$_n(4000)$ & $z$, $\log \mu_*$                & $-0.24$ & $<0.01$ & $\rm{sSFR_{fiber}}>10^{-11.3}\,yr^{-1}$\\
$[9]$&$\log\,s_{\rm arm}$ & D$_n(4000)$ & $z$, $C$                         & $-0.26$ & $<0.01$ & $\rm{sSFR_{fiber}}>10^{-11.3}\,yr^{-1}$\\
$[10]$&$\log\,s_{\rm arm}$ & D$_n(4000)$ & $z$, $\log R_{50}$, $\log M_*$, $\log \mu_*$, $C$ & $-0.38$ & $<0.01$ & $\rm{sSFR_{fiber}}>10^{-11.3}\,yr^{-1}$\\
\hline
$[11]$&$\log\,s_{\rm arm}$ & H$\delta_{\rm A}$ & $\dots$                          & 0.20 & $<0.01$ & $\rm{sSFR_{fiber}}>10^{-11.3}\,yr^{-1}$\\
$[12]$&$\log\,s_{\rm arm}$ & H$\delta_{\rm A}$ & $z$, $\log M_*$                  & 0.26 & $<0.01$ & $\rm{sSFR_{fiber}}>10^{-11.3}\,yr^{-1}$\\
$[13]$&$\log\,s_{\rm arm}$ & H$\delta_{\rm A}$ & $z$, $\log \mu_*$                & 0.22 & $<0.01$ & $\rm{sSFR_{fiber}}>10^{-11.3}\,yr^{-1}$\\
$[14]$&$\log\,s_{\rm arm}$ & H$\delta_{\rm A}$ & $z$, $C$                         & 0.24 & $<0.01$ & $\rm{sSFR_{fiber}}>10^{-11.3}\,yr^{-1}$\\
$[15]$&$\log\,s_{\rm arm}$ & H$\delta_{\rm A}$ & $z$, $\log R_{50}$, $\log M_*$, $\log \mu_*$, $C$ & 0.35 & $<0.01$ & $\rm{sSFR_{fiber}}>10^{-11.3}\,yr^{-1}$\\
\hline
$[16]$&$\log\,s_{\rm arm}$ &  $C({\rm sSFR})$ & $\dots$                          & 0.19 & $<0.01$ & $\rm{sSFR_{fiber}}>10^{-11.3}\,yr^{-1}$\\
$[17]$&$\log\,s_{\rm arm}$ &  $C({\rm sSFR})$ & $z$, $\log M_*$                  & 0.19 &$<0.01$ & $\rm{sSFR_{fiber}}>10^{-11.3}\,yr^{-1}$\\
$[18]$&$\log\,s_{\rm arm}$ &  $C({\rm sSFR})$ & $z$, $\log \mu_*$                & 0.27 &$<0.01$ & $\rm{sSFR_{fiber}}>10^{-11.3}\,yr^{-1}$\\
$[19]$&$\log\,s_{\rm arm}$ &  $C({\rm sSFR})$ & $z$, $C$                         & 0.20 &$<0.01$ & $\rm{sSFR_{fiber}}>10^{-11.3}\,yr^{-1}$\\
$[20]$&$\log\,s_{\rm arm}$ &  $C({\rm sSFR})$ & $z$, $\log R_{50}$, $\log M_*$, $\log \mu_*$, $C$ & 0.26 &$<0.01$ & $\rm{sSFR_{fiber}}>10^{-11.3}\,yr^{-1}$ \\
\hline
$[21]$&$\log\,s_{\rm arm}$ & H$\delta_{\rm A}$ &  D$_n(4000)$      & 0.02 & 0.45 & $\rm{sSFR_{fiber}}>10^{-11.3}\,yr^{-1}$\\
$[22]$&$\log\,s_{\rm arm}$ &  $z$ & $\log M_*$, $\log \mu_*$, $C$  & 0.03 & 0.07 & $\rm{sSFR_{fiber}}>10^{-11.3}\,yr^{-1}$\\
\hline
$[23]$&$\log\,s_{\rm arm}$ & $\log \Delta\Sigma_1$ & $\dots$         & $-0.30$ & $<0.01$ & Galaxies with available $\Delta\Sigma_1$ \\
$[24]$&$\log\,s_{\rm arm}$ & $C$ & $\dots$                           & $-0.23$ & $<0.01$ & Galaxies with available $\Delta\Sigma_1$ \\
$[25]$&$\log\,s_{\rm arm}$ & $\log \Delta\Sigma_1$ & $\log M_*$, $C$ & $-0.19$ & $<0.01$ & Galaxies with available $\Delta\Sigma_1$ \\
$[26]$&$\log\,s_{\rm arm}$ & $\log \Delta\Sigma_1$ & $\dots$         & $-0.06$ & $0.08$ & $\log \Delta\Sigma_1 < 0$\\
$[27]$&$\log\,s_{\rm arm}$ & $\log \Delta\Sigma_1$ & $\dots$         & $-0.17$ & $<0.01$ & $\log \Delta\Sigma_1 \geq 0$\\
$[28]$&$\log\,s_{\rm arm}$ & $C$ & $\log \Delta\Sigma_1$, $\log M_*$ & $0.11$ & $<0.01$ & $\log \Delta\Sigma_1 < 0$\\
$[29]$&$\log\,s_{\rm arm}$ & $C$ & $\log \Delta\Sigma_1$, $\log M_*$ & $-0.08$ & $0.03$ & $\log \Delta\Sigma_1 \geq 0$\\
\hline
\hline
\end{tabular}
\tablefoot{Column (1): row number. Column (2): spiral arm strength, the first parameter for computing correlation coefficient. Column (3): the second parameter for computing correlation coefficient. Column (4): mutual dependence to be removed. Column (5): partial correlation coefficient. Column (6): the $p$ value for testing non-correlation. Column (7): sample for the calculation. 
}
\label{tb1}
\end{table*}

\section{Discussion}\label{discussion}

\subsection{Gas inflow driven by spiral arms}

Secular evolution triggered by disk instabilities is essential to explain the growth of pseudo bulges. Bars and spirals are the most common non-axisymmetric structures in disk galaxies. Bar-driven instability plays a vital role \citep{Kormendy2004}. Bars drive the gas in the galactic disk, outward to form a ring and inward to the galaxy centers \citep{Athanassoula1992b, Sellwood1993, Patsis2000, Regan2004, Combes2008, Haan2009, KimSeoKim2012, Combes2014, Sormani2015, Prieto2005}. Many observational studies have reported the gas inflow caused by bars and the associated enhanced central star formation \citep[e.g.,][]{Sheth2005, Regan2006, Wang2012, Lin2017, Chown2019, Wang2020, Simon2020, Simon2021}. In particular, \cite{Wang2012, Wang2020} and  used a ratio of fiber sSFR to global sSFR to study the bar effect and found enhancement of central star formation in strongly barred galaxies.

Spiral-driven instabilities are involved in the secular evolution processes. Both in theory \citep{Roberts1969, Kalnajs1972, Roberts1972, Lubow1986, Hopkins2011} and simulations \citep{Kim2014, Kim3_2014, Kim2020, Baba2016}, spiral arms can induce a shock on gas clouds and drive gas clouds inflow. Observational evidence supporting the spiral-shock picture has been found. By studying the molecular gas surface density contrasts of 67 star-forming galaxies in the PHANGS-ALMA CO (2–1) survey, \cite{Meidt2021} find that the logarithmic CO contrasts on 150 pc scales are higher than the logarithmic 3.6\,$\mu$m contrasts in a correlation steeper than linear even in the presence of weak or flocculent spiral arms, in agreement with the compression of gas by shocks.  The spiral shock could also explain the high number density and mass in the mass spectrum of gas clouds along the arms \citep{Colombo2014}, shorter gas depletion associated with arms \cite[][but see Foyle et al. 2010]{Rebolledo2012}, enhanced global sSFR in strongly-armed galaxies \citep{Seigar2002, Kendall2015, Yu2021}, offset in pitch angle of different tracers \citep{YuHo2018, MartinezGarcia2014, Egusa2009}, although the turbulence and streaming motions in the dense gas reservoir prevent cloud collapse and curtail star formation efficiency \citep{Meidt2013, Leroy2017}. The relative position, morphology, and kinematics of gaseous and stellar mass in the Milky Way are consistent with models based on the spiral shock \citep{Sakai2015, Hao2021}. Signatures supporting the inflow of gas driven by spiral arms have been detected, although a small sample size limits these studies. \cite{Regan2006} found two unbarred spiral galaxies out of six galaxies having central excess in the 8\,$\mu$m and CO emission above the inward extrapolation of an exponential disk. Some unbarred galaxies could have a high concentration of gas distribution, despite that they are less common than in barred galaxies \citep{Sheth2005, Kuno2007}. The lower frequency for unbarred spiral galaxies exhibiting highly concentrated gas distribution than barred galaxies may result from the stronger bar effect than spiral.  Although simulations comparing the inflow rates driven by bars and spirals are lacking, we may get some clues from the statistics of bar and spiral strength. By utilizing arm/inter-arm contrast, gravitational torque, or Fourier amplitude as a measure of the strength of bars and spirals, it has been shown that bars in barred galaxies are averagely stronger than spirals in unbarred galaxies \citep{Buta2005, Durbala2009, Bittner2017}. The relative weakness of the spiral effect may also be reflected by the moderate correlation coefficients ($\sim$\,0.2 to 0.28) as seen in Figure~\ref{fib_s} and \ref{longsf}. Although the arms may be weaker than bars and they occupy different radial regions, it is possible that in barred galaxies, arms firstly deliver the gas to the radial range within the bar, which successively drives the gas flow toward the center. This join effect was highlighted in \cite{Wang2020}. Our results suggest that the spiral effect is statistically significant and is thus indispensable for understanding the galaxy secular evolution.  It would be worthwhile to investigate in the future, with a large sample, how the properties of spiral arms influence the radial distribution of cold gas.

\subsection{Implication on secular growth of pseudo bulges}

The subsequent star formation followed by gas inflow driven by disk instabilities leads to the growth of central pseudo bulges \citep{Kormendy2004}. We have shown in Section~\ref{results} that spiral arms enhance the central star formation rate of both short and long timescale in a continuous manner, implying spiral arms may also play a role in the secular growth of pseudo bulges.

Recently, \cite{Luo2020} study the relative central stellar-mass surface density within 1\,kpc ($\Delta \Sigma_1$) and found that classical bulges have high $\Delta \Sigma_1$ ($\log\Sigma_1 \geq 0$), while pseudo bulges have low $\Delta \Sigma_1$ ($\log\Sigma_1 < 0$). This method to classify bulge types is in line with that based on the Kormendy relation.  In order to investigate the spiral effect, we cross-match our parent sample (including both centrally star-forming and centrally quenched SFMS galaxies) with the sample in \cite{Luo2020} and found 1738 objects in common. Figure~\ref{sigma1} compares the $\Delta\Sigma_1$ with arm strength. There is a moderate trend that galaxies with stronger spiral arms tend to have lower $\Delta\Sigma_1$ with Pearson correlation coefficient $\rho=-0.30$ and $p$ value $<0.01$ (row [23] in Table~\ref{tb1}). It suggests that a galaxy with strong spiral arms tend to have a pseudo bulge in the centers.

The galaxy concentration index $C$ directly reflects the global shape of the surface brightness profile. The larger the bulge relative to the disk (higher bulge fraction), the more prominent the central profile, and the higher the $C$. The quantity $\Delta\Sigma_1$ measures bulge density (thus classical versus pseudo bulges) rather than the bulge fraction, despite the existence of a relation between $\Delta\Sigma_1$ and bulge fraction. The  light fraction of a pseudo bulge could be larger than that of a classical bulge \citep{Gadotti2009, Gao2020}.

The connection between arm strength and $\Delta\Sigma_1$ arises partly due to the classical bulge if it weakens the arms. In fact, a larger classical bulge will decrease the mass fraction of dynamically active disk to suppress the spiral arms \citep{Bertin1989}, resulting in a trend of increasing concentration with weakening spiral arms \citep{YuHo2020}. We find a similar trend with $\rho=-0.23$ and $p<0.01$ (row [24] in Table~\ref{tb1}). Meanwhile, classical bulges tend to reside in more massive galaxies \citep{Luo2020}.  When studying the actual dependence of pseudo bulges on spiral arm strength, one need to control the suppression effect on spiral arms caused by the classical bulge, the significance of which is properly indicated by its bulge fraction (concentration index) and the galaxy stellar mass.  We thus calculate the partial correlation coefficient between arm strength and $\Delta \Sigma_1$ by removing their mutual dependence on $\log M_*$ and $C$, which yields $\rho=-0.19$ with $p$ value $<0.01$ (row [25] in Table~\ref{tb1}). The arm strength-$\Delta\Sigma_1$ relation become weaker but remains significant after the $\log M_*$ and $C$ are controlled, and thus this relation is likely in part driven by suppression of the arms by classical bulges and in part by central star formation triggered by spiral arms.

When the sample is regrouped into pseudo bulges ($\log\,\Delta \Sigma_1 < 0$) and classical bulges ($\log\,\Delta \Sigma_1 \geq 0$), the correlations between $\Delta \Sigma_1$ and arm strength become much shallower ($\rho=-0.06$, $p=0.08>0.05$ for pseudo bulges; $\rho=-0.17$, $p<0.01$ for classical bulges.), perhaps because of the small dynamic range in $\Delta \Sigma_1$ in each subgroup. However, in the subgroup of galaxies hosting pseudo bulges, we detect a positive concentration-arm strength relation (row [28] in Table~\ref{tb1}), rather than the suppression of spirals by classical bulges.  In galaxies with pseudo bulges, stronger spiral arms tend to have higher galaxy concentration ($\rho=0.11$, $p<0.01$), for a given $\log\,\Delta \Sigma_1$ and $\log\,M_*$. The concentration of galaxies hosting pseudo bulges is possibly elevated by the larger pseudo bulges, as classical bulges are not included. Consistent with the arm suppression, galaxies hosting classical bulges present a weak inverse correlation with $\rho=-0.08$ and $p=0.03$ (row [29] in Table~1). In Figure~\ref{CBPB}, we plot the distribution of $C$ for the strong-armed sample and control sample, which have similar $\Delta \Sigma_1$ and $M_*$ ($\lvert\Delta\log M_{\rm{*}}\rvert\leq 0.1$ \& $\lvert\Delta\log\Sigma_1\rvert\leq 0.1$). For galaxies with pseudo bulges, the strong-armed sample on average has 0.09 higher $C$ than the control weak-armed sample. For those with classical bulges, the $C$ of the strong-armed sample is on average 0.12 lower, although the $p$-value is greater than 0.05. Our results suggest that spiral arms may help build and grow the pseudo bulges.

To shed more light on how spiral arms and bulge types influence each other, we follow the strategy in \cite{Luo2020} to classify 798 pseudo bulges ($\log\Delta\Sigma_1<0$ \& D$_n(4000)<1.6$), 369 star-forming classical bulges ($\log\Delta\Sigma_1\geq 0$ \& D$_n(4000)<1.6$), and 429 quenched classical bulges ($\log\Delta\Sigma_1\geq 0$ \& D$_n(4000)\geq 1.6$). Figure~\ref{mc} presents the $C$ against $M_*$ for the three bulge types. Galaxies with pseudo bulges tend to be less massive and less concentrated. Galaxies with star-forming classical bulges span a wide range. Galaxies with quenched classical bulges tend to have higher mass and higher concentration. In Figure~\ref{histbul}, we illustrate the distribution of arm strength for galaxies hosting pseudo bulges, marked in blue, classical bulges, marked in black, star-forming classical bulges, marked in green, and quenched classical bulges, marked in red. Spiral arms associated with a pseudo bulge tend to be stronger than those with a classical bulge. In the classical bulges population, galaxies with star-forming classical bulges tend to have stronger arms than with quenched classical bulges. Similar behavior can be seen in Figure~\ref{frac}, the fraction of galaxies hosting pseudo bulges increases with increasing arm strength, while the fraction of classical bulges decreases with increasing arm strength. In the classical bulges population, the fraction of galaxies hosting quenched classical bulges decreases with increasing arm strength, while the fraction of star-forming classical bulges remains relatively unchanged. Classical bulges associated with strong spiral arms ($2+\log s_{\rm arm}>1.35$) tend to be star-forming.

Classical bulges form through rapid processes of violent relaxation or gaseous dissipation at an early epoch. A scenario is that major mergers provide violent relaxation and drive rapid gas inflow to trigger central starbursts to create highly concentrated bulge with stars of random motion \citep{Hopkins2009a, Hopkins2009b, Hopkins2010, Brooks2016, Tonini2016, RodriguezGomez2017}. Minor mergers play a lesser role in the formation or growth of classical bulges \citep{Aguerri2001, ElicheMoral2006, Hopkins2010}. A second scenario is that gas-rich disks at high redshift are highly turbulent and have giant star-forming clumps formed by gravitational instabilities \citep{Elmegreen2005}. The bound clumps interact, lose angular momentum, and migrate to the center to form a classical bulge \citep{Noguchi1999, Bournaud2007, Bournaud2009, Elmegreen2008, Bournaud2016}. The coalescence of massive disk clumps has a similar behavior of orbital mixing to a major merger. 

At later stages, there are enough hot stars in a thick disk and bulges arising from the previous stages so that gravitational instabilities produce spirals rather than clumps \citep{Bournaud2009}. The spiral structure in the disks then occurs at $1.4 \lesssim z \lesssim 1.8$, when disks settle down with rotation motion dominated over turbulent motions in the gas and massive clumps become less frequent \citep{Elmegreen2014}. The onset of spiral arms follows a morphological transformation sequence from clumps to ``woolly arms'', to irregular long arms, and finally to normal spiral structure \citep{Elmegreen2014}. The pre-existing classical bulges and their associated hot thick disks influence the development of spiral structure through reducing the mass fraction of the dynamically active disk that reacts to spiral perturbation \citep{Bertin1989}. Thus the galaxies with prominent classical bulges have weak spiral arms \citep{YuHo2020}. A fraction of massive star-forming galaxies has started inside-out quenching at $z \sim 2$ \citep{Tacchella2015, Tacchella2018}. In the nearby universe, more massive galaxies exhibit a greater fraction of inside-out quenching compared with low massive ones in all environments, which may be explained by the morphological quenching \citep{Lin2019}. Perhaps consistent with the inside-out quenching, galaxies with quenched classical bulges have a higher mass and higher concentration than star-forming classical bulges (Figure~\ref{mc}). The influence of classical bulge on spiral arms in part explains the correlation between bulge types and arm strength, but not entirely since a residual interdependence between arm strength and bulge types indicator after removing effects of concentration and mass was detected (row [24] in Table~\ref{tb1}).

Bars appear at about $z\sim 1$ \citep{Sheth2008} and drive secular evolution \citep{Kormendy2004}. The two-dimensional multiple-components decomposition shows that the bulges of barred galaxies do not have a different Kormendy relation than unbarred galaxies \citep{Gao2020}. Likewise, there is no difference in the relationships between relative central surface density and other global galaxy properties for barred and unbarred galaxies \citep{Luo2020}.  One possible explanation is that the bar is short-lived \citep{Bournaud2002} instead of long-lived \citep{Athanassoula2013}. The compact classical bulges or central black holes may weaken and even destruct bars \citep{Combes1996, Bournaud2005}, which can re-form through gas accretion \citep{Combes1996, Block2002, Bournaud2002, Bournaud2005}. Some of the unbarred galaxies observed in our sample may have previously hosted a bar, which facilitates the formation of pseudo bulges during the bar's lifetime.  Alternatively, most of the bars are long-lived, but in unbarred galaxies, non-axisymmetric structures other than bars such as spirals also drive gas inflow and participate in the build-up of pseudo bulges.

When disks have settled down and normal spiral arms have developed, the spiral arms drive gas flow toward the center primarily by dissipation of angular momentum at spiral shocks, secondarily by gravitational torque of the spiral potential \citep{Kalnajs1972, Roberts1972, Lubow1986, Hopkins2011, Kim2014, Kim2020}, and by self-gravitational torque of the gaseous component in a minor way \citep{Kim2014}. Stronger arms trigger a higher mass inflow rate of gas \citep{Kim2014}. The inflow gas feeds the central star formation so that stronger spiral arms have a higher central star formation rate even after the effects of stellar mass, surface density, concentration, and redshift have been removed (Section~\ref{results}). The gas inflow is not rapid enough to trigger a central burst of star formation (Section~\ref{csf}). The subsequent star formation contributes to the build-up of the pseudo bulges, resulting in a connection between relative central stellar surface density and spiral arm strength, irrespective of galaxy mass and concentration (Figure~\ref{sigma1}).

In the same vein, spiral arms may enhance central star formation in some less massive galaxies with a pre-existing small classical bulge (green histogram in Figure~\ref{mc}). Compared with their massive counterparts (red histogram in Figure~\ref{mc}), less massive galaxies could have a substantial gas reservoir since the gas fraction increases with lower stellar mass \citep{Saintonge2017}. Spiral arms are still present, possibly maintained by the cold gas \citep{Bertin1989, Yu2021}.  The arms drive gas to funnel to the center.  Assuming that different types of bulges can co-exist \citep{Athanassoula2005}, the newly inflowing gas onto a pre-existing classical bulge should maintain strong in-plane rotational dynamics and, therefore form a disky pseudo bulge in addition to the classical bulge. Evidence for the coexistence of classical bulges and disky pseudo bulges has been reported \citep{Erwin2015}. Galaxies with a star-forming classical bulge likely host an additional pseudo bulge, a hypothesis that may be tested in the future. However, this picture hardly applies to massive galaxies with a prominent classical bulge, as the inner gaseous disk, if present, is likely to be stabilized against star formation by the prominent classical bulge \citep{Martig2009}.

Together with the suppression of spiral arm strength by classical bulges, spiral-driven secular evolution leads to the relationship between bulge types and spiral arms (Figure~\ref{histbul} and \ref{frac}).

\section{Conclusions}\label{conclusions}

We have used 2779 nearby relatively face-on unbarred star-forming main-sequence (SFMS) spiral galaxies, derived from the SDSS, to investigate the hypothesis that spiral-driven instabilities drive gas inward to enhance galaxy central star formation.  Galactic bars can trigger a central star formation and then contribute to the secular growth of galaxy centers \citep{Kormendy2004}, but less is known about the spiral effect. The star formation properties in the central 1--3\,kpc region were derived from the SDSS spectra. Specifically, we utilize ${\rm sSFR_{fiber}}$, computed based on emission lines, from the MPA-JHU catalog to trace central on-going star formation averaged over the pass $\sim$\,10\,Myr, H$\delta_{\rm A}$ to probe star formation on intermediate timescales of 0.1--1\,Gyr prior to observation, and D$_n(4000)$ to indicate the luminosity-weighted mean stellar age of longer timescales of several Gyr. A ratio of ${\rm sSFR_{fiber}}$ to ${\rm sSFR_{global}}$, $C({\rm sSFR})= {\rm \log(sSFR_{fiber}/sSFR_{\rm global})}$, where ${\rm sSFR_{global}}$ is measured for the whole galaxy acquired from \cite{Salim2018}, is employed as a measure of enhancement of central star formation relative to global star formation.  We are essentially relating the spiral arms occupying the extended optical disk to the central most region of the galaxy. The 2779 SFMS spiral galaxies are further separated into two subsamples of 2056 centrally star-forming SFMS galaxies and 723 centrally quenched SFMS galaxies. To avoid possible influence by quenching processes, only the centrally star-forming SFMS galaxies are used when studying the impact of spiral arms, but both of them are used when studying the bulge types.

The relative amplitude of spiral arms ($s_{\rm arm}$) is defined as the mean Fourier amplitude relative to an axisymmetric disk over the disk region. The logarithmic form of the relative amplitude ($2+\log s_{\rm arm}$) is used as a measure of the strength of spiral arms. Biases caused by noise are corrected.  We investigate the impact of spiral arms on central star formation by comparing spiral arm strength with central star formation properties. We also isolate the effect of spiral arms by removing effects of redshift ($z$), stellar mass ($\log M_{*}$), stellar surface density ($\log \mu_*$), and concentration ($C$). Our main findings are as follows.

\begin{enumerate}
  \item Galaxies with stronger spiral arms not only tend to have more intense central sSFR, larger H$\delta_{\rm A}$, and lower D$_{n}(4000)$, but also have enhanced $C({\rm sSFR})$. \\

  \item Compared with weak-armed galaxies of similar $z$, $\log M_{*}$, $\log \mu_*$, and $C$, the central star formation enhancement in strong-armed galaxies is still significant. It is further verified by partial correlation coefficients. These results suggest that spiral arms can enhance central star formation and it is a true effect. \\
 
  \item The central starburst galaxies have both weak and strong arms and only $\sim$\,10\% of galaxies with very strong spiral arms (top 5\% of arm strength) have central starburst. Likewise, there is no apparent excess in H$\delta_{\rm A}$ in strong spirals for a given D$_{n}(4000)$. Strong spiral arm is thus not a sufficient or necessary condition to trigger a central starburst. It implies that the spiral-induced central star formation is continuous instead of bursty.\\
 
  \item There is a trend of increasing arm strength with lower relative central stellar surface density, suggesting strong spiral arms tend to have pseudo bulges. Moreover, in galaxies hosting pseudo bulges, stronger spiral arms tend to have higher concentration index, which is possibly elevated by the more prominent pseudo bulges. These results suggest that spiral arms may play a role in the build-up of pseudo bulges.\\
 
  \item Galaxies with increasing spiral arm strength tend to have an increasing fraction of pseudo bulges, a relatively unchanged fraction of star-forming classical bulges, and a decreasing fraction of quenched classical bulges. This relationship is partly attributed to the suppression of spirals by classical bulges \citep{Bertin1989, YuHo2020} and partly to the central star formation driven by spirals, which builds the pseudo bulges. 
\end{enumerate}

We explain our results in a picture where spiral arms transport cold gas inward to trigger a continuous central star formation. The subsequent star formation contributes to the secular growth of pseudo bulges. Spiral arms thus play an essential role in the secular evolution of disk galaxies.

\begin{acknowledgements}
LCH was supported by the  National Science Foundation of China (11721303, 11991052) and the National Key R\&D Program of China (2016YFA0400702). SYU acknowledgements the support from the Alexander von Humboldt Foundation.  We thank the referee for constructive criticism that helped to improve the quality and presentation of the paper. We benefited from discussions with Veselina Kalinova, Dario Colombo, and Karl Menten.  SYU is indebted to Karl Menten for his great support during the pandemic. 
\end{acknowledgements}

\bibliographystyle{aa}


\end{document}